\documentclass[fleqn,usenatbib]{rasti}

\usepackage{newtxtext,newtxmath}

\usepackage[T1]{fontenc}

\DeclareRobustCommand{\VAN}[3]{#2}
\let\VANthebibliography\thebibliography
\def\thebibliography{\DeclareRobustCommand{\VAN}[3]{##3}\VANthebibliography}

\usepackage{graphicx}	
\usepackage{amsmath}
\usepackage{multicol,tabularx,capt-of}
\usepackage{multirow}
\usepackage{comment}
\usepackage[normalem]{ulem}

\title[Contrastive learning in astrophysics]{A brief review of contrastive learning applied to astrophysics}

\author[Huertas-Company, Sarmiento \& Knapen]{
Marc Huertas-Company$^{1,2,3,4,5}$\thanks{E-mail: mhuertas@iac.es},
Regina Sarmiento$^{1,2}$
and Johan H. Knapen$^{1,2}$
\\
$^{1}$Instituto de Astrof\'isica de Canarias (IAC), La Laguna, E-38205, Spain\\
$^{2}$Departamento de Astrof\'isica, Universidad de La Laguna (ULL), E-38200, La Laguna, Spain\\
$^{3}$Observatoire de Paris, LERMA, PSL University, 61 avenue de l'Observatoire, F-75014 Paris, France\\
$^{4}$Universit\'e Paris-Cit\'e, 5 Rue Thomas Mann, 75014 Paris, France \\
$^{5}$Center for Computation Astrophysics, Flatiron Institute, 162 5th avenue, 10015 New York, USA
}

\date{Accepted XXX. Received YYY; in original form ZZZ}

\pubyear{2022}

\begin{document}
\label{firstpage}
\pagerange{\pageref{firstpage}--\pageref{lastpage}}
\maketitle

\begin{abstract}
Reliable tools to extract patterns from high-dimensionality spaces are becoming more necessary as astronomical datasets increase both in volume and complexity. Contrastive Learning is a self-supervised machine learning algorithm that extracts informative measurements from multi-dimensional datasets, which has become increasingly popular in the computer vision and Machine Learning communities in recent years. To do so, it maximizes the agreement between the information extracted from augmented versions of the same input data, making the final representation invariant to the applied transformations. Contrastive Learning is particularly useful in astronomy for removing known instrumental effects and for performing supervised classifications and regressions with a limited amount of available labels, showing a promising avenue towards \emph{Foundation Models}. This short review paper briefly summarizes the main concepts behind contrastive learning and reviews the first promising applications to astronomy. We include some practical recommendations on which applications are particularly attractive for contrastive learning. 

\end{abstract}

\begin{keywords}
methods: data analysis -- methods: statistical -- methods: miscellaneous -- techniques: miscellaneous
\end{keywords}



\section{Introduction}

As astronomical data become larger in volume and higher in dimension, new tools are needed to visualize and extract the relevant information contained in these datasets. Although dating back to the 1950s and 1960s (see, e.g., \citealp{e08667bd2e844956b9d345b9e957634d} for a comprehensive and historical overview), the field of machine learning (ML) has, over the past decade in particular, proven versatile as a statistical tool for data analysis and the deduction and prediction of trends from massive data sets (see, e.g., \citealp{2023PASA...40....1H,2022arXiv221103796S} for recent reviews on deep learning applied to astronomy and astrophysics). Whereas supervised ML is widely used in astronomy for classification and other tasks, it is in many situations limited by the availability of labeled samples. Since most data is generally unlabeled, self- and unsupervised ML are potentially powerful tools for uncovering correlations hidden in complex data sets. Applications in astronomy are still relatively limited, however, mainly because it is generally difficult to interpret the results which can also be biased by non-physical properties of the data. In this paper, we review the use and promise of contrastive learning (CL) in astrophysics. CL is a self-supervised representation learning technique that aims to combine the power of unsupervised ML while avoiding some of its most obvious dangers. {This review work assumes that the reader is familiar with basic concepts of ML and, in particular, with modern deep learning techniques.

\subsection{A brief history of representation learning}

\textit{Representation learning} (e.g., \citealp{bengio2013representation}) refers to the general idea of automatically learning a mapping between raw high-dimensional data and a feature space—typically but not always of smaller dimension —that efficiently captures the relevant and most informative correlations in the data. The concept of representation learning is tightly connected to those of \textit{dimensionality reduction} and \textit{feature extraction}, although some subtle differences exist. Feature extraction, which is the process of extracting meaningful information from data, can be manual or automatic while representation learning generally refers to techniques with no direct human supervision. \textit{Dimensionality reduction} methods (e.g., \citealp{van2009dimensionality}) also find a lower-dimension representation of the data but do not necessarily offer a mapping that can be used to evaluate new data points, as opposed to representation learning. The textbooks by \cite{bishop2006pattern} and~\cite{pml1Book} provide excellent introductions to these basic concepts.

The origins of representation learning go back to principal component analysis (PCA, \citealp{PCA-1901}), where a high-dimensional space can be represented by a reduced number of orthogonal eigenvectors. With the goal of mapping high-dimensional data onto a lower-dimensional space, algorithms that preserve distances like multidimensional scaling (MDS, \citealp{MDS}) were developed. Like PCA, MDS is a robust linear approach to extract features but is based on pairwise distances. While PCA finds linearly uncorrelated parameters that minimize the variance in the input data, MDS finds a linear decomposition that best reproduces the pairwise distances of the input space. However, these methods have poor performance when the data have a nonlinear distribution as they provide a linear decomposition of the data. This motivated the development of methods that bypass the non-linearity of the data with an additional transformation (Kernel PCA, \citealt{KernelPCA}) or that also (or only) consider local distances (neighbours in the input space will also be neighbours in the feature space) like isomap (\citealt{isomap}), locally linear embedding \citep{LLE}, Laplacian eigenmaps \citep{Belkin:2003}, Hessian eigenmaps \citep{Hess-maps} or t-SNE \citep{vanDerMaaten2008}. A limitation of these latter approaches is that they are unable to predict the projection of a new input object into the lower dimension space which makes them suboptimal for representation learning. More recent tools overcome this (e.g., UMAP, \citealt{umap}).

\subsection{Representation learning in the deep learning era}

With the rapid advances in neural network architectures over the past decade, deep learning-based methods for representation learning have become common. The general idea is to use a neural network to approximate some properties of a dataset~(e.g., \citealp{hinton2006reducing}).
In the process, the neural network is expected to learn some meaningful representations of the data. There are generally two broad types of approaches which are referred to as \textit{generative} and \textit{discriminative}. They differ based on the target function the neural network is used to approximate.

Given a dataset $X$ of high dimension - typically images or spectra in astronomy - and eventually some labels $Y$ associated with it - class or physical quantity - generative models aim at estimating the probability distribution of the data $\{x\in X\}$ $p(x)$ - or $p(x|y)$ if some labels $\{y\in Y\}$ are available - by using a latent variable $z$ of generally lower dimension than $x$. For example, $p(x)$ can be the joint probability distribution of the pixel values of a set of images~\footnote{In practice, a generative model learns $p(x|z)$ which can then be used to approximate $p(x)$.}. Once trained, $p(x)$ can be sampled to generate new data points. However, for the purpose of representation learning, the interesting part is that in the process of learning $p(x)$, the network encapsulates some information about the dataset in the latent variable $z$, which can be, to some extent, interpreted as a representation of $x$.

Discriminative methods, on the other hand, estimate a conditional probability distribution $p(y|x)$. As opposed to generative approaches, which can be applied with or without labels, they require a label by construction. The neural network is indeed used to approximate a non-linear mapping between $x$ and some label $y$.  As for the generative case, in the process of learning the mapping a (lower) dimensionality projection of the data is represented in the layers of the network, which can be used as a representation space.

Discriminative methods are usually trained with supervised approaches since they require labels. In fact, any modern neural network trained for classification or regression can be used as a representation learning framework (see, for example, \citealp{Walmsley-2022-1} for an application to galaxy morphology). Generative approaches can be trained both in supervised and unsupervised mode. The most popular approaches in the astronomy literature over the past years are Variational AutoEncoders \citep{kingma2013auto,rezende2014stochastic,doersch2016tutorial,higgins2017beta,chen2018isolating} or Generative Adversarial Networks \citep{goodfellow2014generative,radford2015unsupervised,salimans2016improved,arjovsky2017wasserstein,karras2019style,brock2018large}. Other generative models such as Neural Flows \citep{rezende2015variational,dinh2016density,kingma2016improved,papamakarios2017masked,grathwohl2018ffjord,chen2018neural} or Diffusion Models \citep{sohl2015deep,song2019generative,ho2020denoising,grathwohl2021your,chen2021stochastic,liu2021learning} are rapidly increasing in popularity, although they do not necessarily require dimensionality reduction.

Both approaches have pros and cons. Discriminative models are usually easy to train but require labels, which limits their applicability. Additionally, the representations learned by these methods are limited by the generality of the labels used for training. For instance, a network trained to identify foreground stars in galaxy images may not produce relevant features for studying galaxy morphology. However, \cite{Walmsley-2022-1} showed that a model trained with a combination of labels describing galaxy morphology generalizes well to new tasks. Generative models attempt to overcome some of these issues, but at an important computational cost. Properly modeling $p(x)$ solely for obtaining representations may be considered overkill and a waste of resources and time.

\subsection{Self-supervised learning}

Self-supervised approaches try to get the best of both worlds – discriminative and generative – by adapting discriminative approaches to the case where no labels are available. They do so by creating a pretext task $\hat{y}$, which the neural network is trained to predict. By using this trick, the neural network can be trained under a discriminative setting without the need for labels. Pretext tasks can be of different types; for example, one can predict the rotation angle of a given image. Table~\ref{table:representation} concisely summarizes the different approaches to representation learning with neural networks and the role of self-supervised learning. In recent years, however, arguably the most successful self-supervised approaches are the so-called contrastive models, whose origins trace back to DrLIM (\citealt{Chopra-2005}, \citealt{Hadsell-2006}). In contrastive models, the pretext task is set to be a measurement of similarity between data points – see Section~\ref{sec:contrastive} for a detailed explanation.  In particular, contrastive models started to attract attention a few years ago when the representations learned through contrastive learning used in a supervised classification problem, achieved better accuracy than a pure supervised training~\citep{SimCLR1}. Since then, there have been numerous applications and proposed improvements in the ML literature. For example~\cite{Le_Khac-2020} give a nice overview of self-supervised and contrastive learning from a pure ML perspective.\\

\begin{table*}
\caption{Approaches to representation learning with neural networks}             
\label{table:representation}      
\centering                          
\begin{tabular}{{|p{1.5cm}|c|c|c|c|}}  
\hline\hline 
\multirow{2}{1.5cm}{\textbf{Approach Data type}} & \multicolumn{2}{c|}{\textbf{Discriminative}} & \multicolumn{2}{c|}{\textbf{Generative}}\\
  \cline{2-5}

& Target Function & Method & Target Function & Method \\
 \hline 
 {\bf Labels} & $p(y|x)$ & All supervised networks with bottleneck & $p(x|y)$ & Conditional generative models \\  
 {\bf No Labels} & $p(\hat{y}|x)$ & Self-supervised learning & $p(x)$ & Generative models, Autoencoders \\  
\hline                                   
\end{tabular}
\end{table*}

Although contrastive learning is a relatively young method, there already exist a number of applications in astrophysics. This short work reviews the main ideas behind self-supervised contrastive learning techniques and how this has been applied to astrophysics so far. We also discuss the potential for future applications from a very practical point of view. \\

The paper proceeds as follows. In Section~\ref{sec:contrastive} we briefly describe the main technical aspects of a contrastive learning framework. Section~\ref{sec:applications} makes a census of the applications of contrastive learning in astronomy so far and Section~\ref{sec:discussion} discusses some practical considerations regarding contrastive learning. The final Section offers a brief conclusion.

\section{What is contrastive learning?}
\label{sec:contrastive}

Contrastive learning (CL) is a self-supervised framework to learn meaningful representations from a dataset $X$. It consists of a trainable function $f:X\rightarrow \mathbb{R}^n$, parametrized by a neural network (NN), that is optimized so that the representations $\boldsymbol{z}=f(\boldsymbol{x})$ - with ${\boldsymbol{z}\in\mathbb{R}^n}$ and ${\boldsymbol{x}\in X}$ - become invariant to different views of the same object.

The different views of the same object are identified as positive pairs (there can be more than one pair per object), while any other combination will represent a negative pair. The key difference with other representation learning methods is that the positive pairs (or neighbours) in CL do not necessarily depend on a distance definition in the input space.

The way the positive pairs are defined determines the pretext task that the NN is optimized to solve. Since the similarity is measured in the representation space, positive pairs can be different data types that represent the same objects (e.g., images and spectra of the same source in astrophysics), a cutout of the input data, or any other transformed copy of it, such that the two versions of each individual input are identified as positive pairs. This is particularly beneficial when complex noise and/or selection effects exist in the input data. In such cases, the individual data points that share semantic information may not be well represented by a linear combination of neighbors in the input space.

Modern CL has been predominantly applied to imaging data, where views are easily defined. For example, Figure~\ref{fig:aug-galaxy_img} illustrates some common image augmentations applied to a galaxy image from the Sloan Digital Sky Survey (SDSS) I DR7~\citep{2009ApJS..182..543A}: cutouts, blurring, rotation (and flip), jitter, color shifts, crop, and the addition of noise.

\begin{figure*}
    \centering
    \includegraphics[width=\hsize]{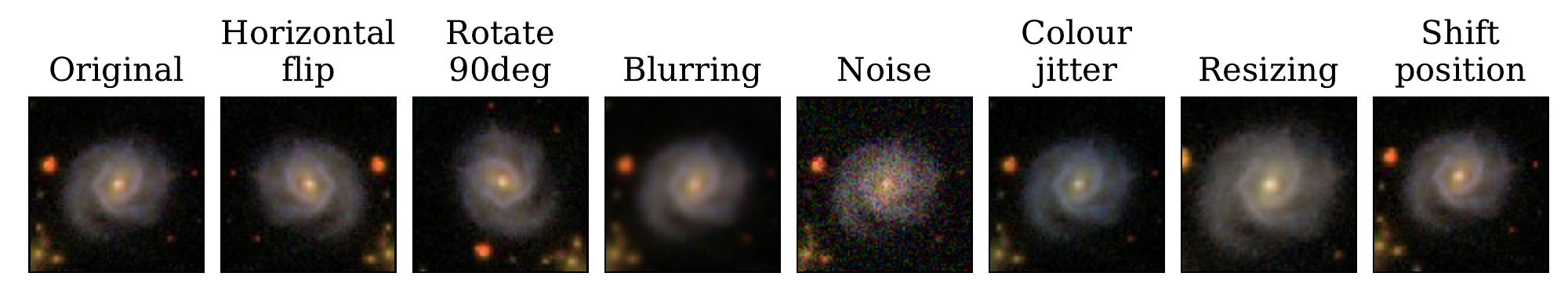}
    \caption{Common transformations used on galaxy imaging to produce positive pairs. From left to right: original $gri$-band composite, horizontal flip, rotation, blurring, noise addition, colour jitter, resizing and position shift.}
    \label{fig:aug-galaxy_img}
\end{figure*}

In the next subsections we present the most commonly used loss functions applied to CL approaches and briefly review the most common NN architectures usually employed. 

\subsection{Contrastive loss}\label{Sec:cont_loss}

All CL frameworks share some version of the so-called contrastive loss function, which is generally optimized such that each positive pair is projected together in the representation space ${\boldsymbol{z}\in\mathbb{R}^n}$ and the negative pairs are repelled, although some of the latest proposed versions do not use negative pairs, as we discuss below.

The loss is calculated in the latent space of representations ${\boldsymbol{z}\in\mathbb{R}^n}$, which is the image of the trainable function $f$, where a pairwise distance metric $\langle\cdot,\cdot\rangle$ is defined. Given a set of representations ${\boldsymbol{z}\in\mathbb{R}^n}$, where each $\boldsymbol{z}$ has one identified positive pair $\boldsymbol{z}^+$ or more, the contrastive loss is a function that is minimized when the distance between positive pairs is minimized. The main difficulty is avoiding the trivial solution where all positive representations collapse into the same representation value. Several approaches have been proposed in the literature, which we briefly review now.

\subsubsection{Spring loss}

\cite{Chopra-2005} and \cite{Hadsell-2006} define a loss function that takes the pairwise Euclidean distance as metric. This resembles a spring system, where positive pairs of $\boldsymbol{z}_i$ are attracted, and the negative pairs that are within an $m$-distance of $\boldsymbol{z}_i$ are repelled. The introduction of negative pairs prevents the collapse into a trivial solution. This loss $L^S$ is defined pairwise for $\boldsymbol{z}_i$ and $\boldsymbol{z}_j$ as

\begin{equation}\label{Eq:springLoss}
L^S_{i,j}= Y\frac{1}{2}\langle \boldsymbol{z}_i, \boldsymbol{z}_j \rangle^2 + \frac{1}{2} (1-Y) \max{0, m-\langle \boldsymbol{z}_i,\boldsymbol{z}_j\rangle},
\end{equation}

where $Y$ is equal to one if $\boldsymbol{z}_i$ and $\boldsymbol{z}_j$ are identified as positive pairs, or zero if not. Therefore, either the left or right term in Eq.,\ref{Eq:springLoss} is canceled, respectively. 

\subsubsection{Triplet loss}

The triplet loss $L^T$ \citep{JMLR-2009}, who use a Mahalanobis distance metric \citep{OASIS-2010} with a bilinear model for the similarity measure, is similar to the spring loss, but each term is calculated considering three representations: the anchor ($i$), a positive ($i^+$), and a negative pair ($n$), as

\begin{equation}
L^T_{i,i^+,n}= \max{0, \langle z_i,z_i^+ \rangle - \langle z_i,z_n \rangle + d} .
\end{equation}

Here, $d$ is a hyperparameter determining the distance between positive and negative pairs. This results in a more relaxed condition on the positive pairs than the Spring loss, as the latter attracts the positive pair to the same point in the representation space while repelling the negatives. In contrast, the Triplet loss only seeks to repel the negative representations a distance $d$ longer than the positive pairs, which favors repelling hard negatives.

\subsubsection{Normalized cross entropy}

Modern contrastive models are based on the noise-contrastive estimation (infoNCE) or normalized temperature-scaled cross entropy (NT-Xent) losses \citep{CPC,Misra-2019,SimCLR1}. It is generally defined element-wise as

\begin{equation}
L^P_{i} = -\log \frac{\exp(\langle \boldsymbol{z}_i, \boldsymbol{z}i^+\rangle/h)} {\sum{k},,\exp(\langle \boldsymbol{z}_i, \boldsymbol{z}_k\rangle/h)},
\label{contrastive_loss}
\end{equation}

for the $i$-th object in the set, where $k$ iterates over all the representations $\boldsymbol{z}$ in the batch, and $\langle,, \rangle$ is a pairwise similarity measure, generally the cosine distance or a bilinear model. $h$ is a normalizing factor - \emph{temperature} - that plays a similar role to $m$ in the previous approach, as it determines the concentration of the representation space by weighing the pairwise distances. This loss can be understood as a multiclass loss, where each representation's correct class corresponds to its positive pair. The probability that a given object is assigned to its class over the other "wrong" classes is parameterized by the softmax function, and a cross-entropy loss is optimized.

As CL frameworks typically benefit from a large number of negative samples, the $L^P$ losses are more efficient (faster convergence) than the $L^S$ ones as every object accounts for the negative pairs of the object $\binom{N}{2}\times N$, while in the $L^S$ and $L^T$, only up to $\binom{N}{2}$ negative pairs are considered in total.

Some works have proposed losses based on mutual information (MI) maximization \citep{hjelm2019learning}, where MI measures the dependence that a random variable $W$ has with another random variable $V$ and is defined as

\begin{equation}
I(w,v) = \mathbb{E}_{p(w, v)} \left[ \log\frac{p(w,v)}{p(w)p(v)} \right].
\end{equation}

While the MI-based approaches were proposed independently, the infoNCE optimization is analogous to the MI one, as the first is equivalent to maximizing a lower bound on MI (\citealt{CPC}).

\subsubsection{Contrastive loss without negative pairs}
More recently, \cite{BYOL} proposed  a loss that is calculated with the cosine distance between the positive pairs only and therefore departs from the standard CL setup:

\begin{equation}
    L^B_{i} = 2-2 \frac{\langle \boldsymbol{z}_i, \boldsymbol{z}_i^+\rangle} {||{z}_i||_2.||{z}_i^+||_2}
    \label{byol_loss}
\end{equation}

To avoid collapsing to the trivial solution, they use a twin network where one branch (target) is prevented to update its weights through back-propagation (\citealt{SimSiam}). Instead, the weights of the target branch are a function of those in the symmetric branch (online) which has the same architecture (see Section\,\ref{Sec:frameworks} for more details)

In addition to these common augmentations, one of the most attractive properties of CL is the possibility to define custom augmentations which are domain-specific and allow one to marginalize over known nuisance effects. We will discuss this in more detail in sections~\ref{sec:applications} and~\ref{sec:discussion}.

\subsection{Contrastive learning architectures}\label{Sec:frameworks}

The representations $\boldsymbol{z}$ are computed using NNs. Over the past years, a number of frameworks have been developed, varying the structure and hyper-parameters of the NNs to obtain better performances.

Given that CL is primarily designed for datasets without labels, the evaluation and comparison of different approaches is not always straightforward. Architectures are therefore generally tested on standardized datasets for which labels do exist. The underlying idea to validate a new approach is that if extracted representations are meaningful, then a simple supervised classifier should be able to correctly classify the data. This way, the performance of the algorithms can be tested on supervised tasks, also known as downstream tasks. Therefore, when an approach is presented as more accurate than another in the literature, it generally refers to the accuracy in the downstream supervised task.

Although it is outside the scope of this brief overview to cover all existing implementations of CL, we attempt to summarize the major architecture conceptions that have resulted in significant performance improvements in downstream tasks in the following. A schematic view of these different architectures is presented in Figure~\ref{fig:frameworks}.

    \subsubsection{ CMC (Contrastive Multiview Coding) }
    \cite{Hadsell-2006} stated that the representations benefit from a large number of negative examples. However, this comes at the computational cost of calculating an increasing number of loss terms and more representations in each learning step. This motivated following works to include a memory bank (\citealt{Misra-2019}, \citealt{Wu-2018}, \citealt{CMC}) that would store the representations calculated in previous iterations to increase the number of negative examples. In each iteration the loss is calculated considering the newly calculated representations in the mini-batch together with $m$ randomly sampled representations from the memory bank. This approach benefits from an adjustable number of negative examples without needing to compute more representations than those in the mini-batch.
    
    \subsubsection{ MoCo (Momentum Contrastive)}
    Other implementations replace the memory bank with a parallel network that acts as a momentum encoder (MoCo, \citealt{MoCo1}), whose weights are updated as a moving average of the main branch's weights. The momentum encoder calculates updated representations, which are later queued in a dictionary (a subset of the training set). The oldest representations in the dictionary are replaced by new ones as the network is trained. In contrast to the memory bank approach, the negative examples are sampled from the dictionary which is progressively updated. This prevents the network from using outdated representations from previous epochs as it is restricted to the representations of the immediately previous mini-batches as negative examples. \cite{MoCo1} found that this framework benefits from a slowly evolving momentum encoder to provide consistent representations over mini-batches and hypothesized that this is due to the rapidly changing main branch encoder.

    
  \subsubsection{SimCLR (Simple Contrastive Learning)}
\cite{SimCLR1} propose a single Convolutional Neural Network (CNN) and no memory bank, so the negative examples are limited by the batch size. This sets a requirement on a minimum amount of memory for training. They show that this setup gets comparable results to the MoCo framework using large batch sizes (1024). \cite{SimCLR1} also prove that more parameters in the network, and more and stronger augmentations result in more accurate representations learned. Another important result of this work is that adding a set of fully connected layers before the loss calculation yields better representations in previous layers. This feature is then included in subsequent works (\citealt{MoCo2}, \citealt{BYOL}).

\subsubsection{BYOL (Bootstrap Your Own Latent)}
In the case of \cite{BYOL} and \cite{SimSiam}, a set of twin networks is implemented. The networks' architectures are identical on both branches except that one branch has an additional projection function. This is the branch where the back-propagation is implemented. The differences between the two implementations shed some light on understanding how these setups work. While BYOL's target branch is progressively updated with the twin network's weights to stabilize the output, SimSiam shares the weights in both branches. BYOL incorporates a set of fully connected layers after the CNN in both branches, although \cite{SimSiam} claim that these layers do not improve the representations learned. However, both approaches agree on the need to limit the back-propagation to the online branch to avoid collapsing to a trivial solution, as the loss is calculated only with positive pairs (see subsection~\ref{Sec:cont_loss}).

\subsubsection{CLIP (Contrastive Language-Image Pre-training)}

Recent architectures have started to use different types of networks for feature extraction, which are optimized for different data types. This allows for the combination of, for example, text and images. For example, \cite{2021arXiv210300020R} train language and image encoders with a contrastive learning model to obtain representations of both inputs. They show that introducing descriptive text of an image in the CL setting produces better generalization and improves zero-shot learning.

\begin{figure*}
    \centering
    \includegraphics[width=\hsize]{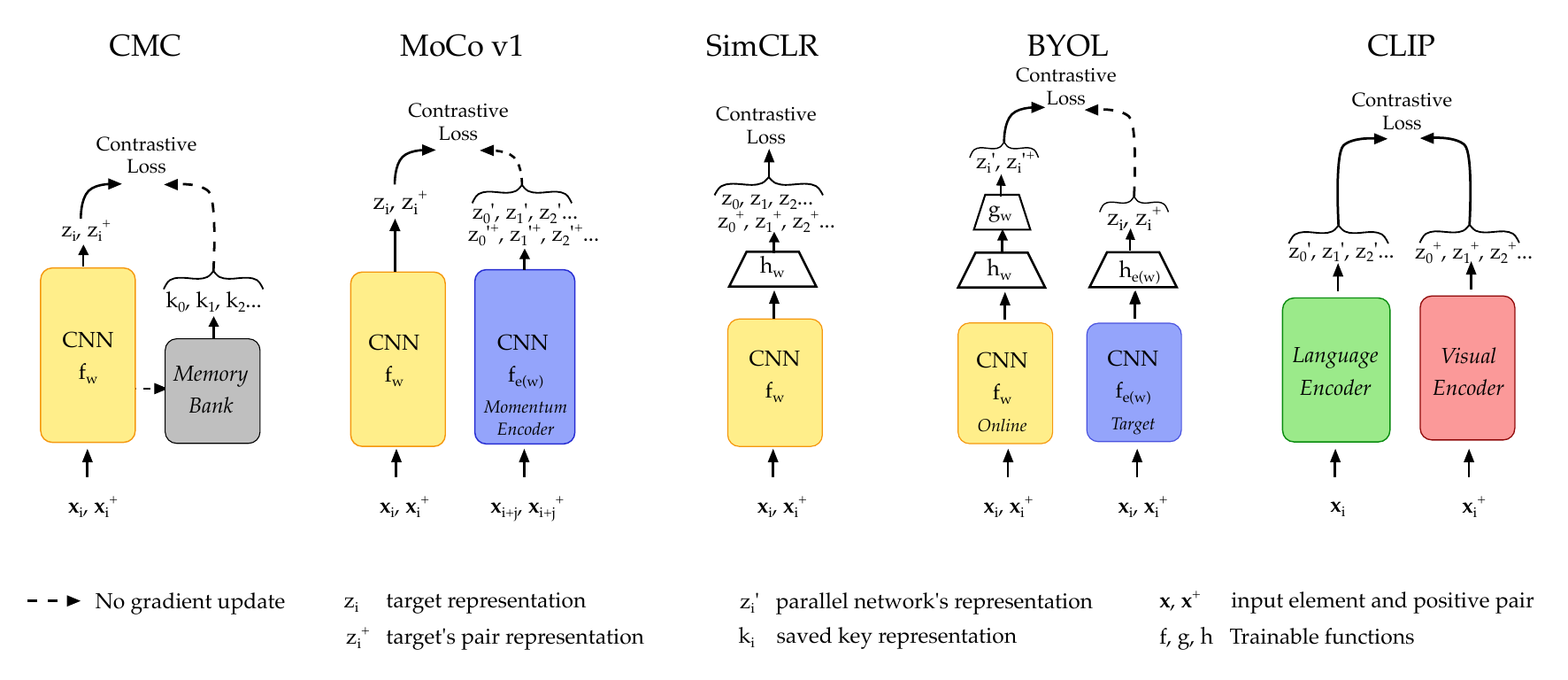}
    \caption{A reduced number of frameworks are shown in a simplified scheme to exemplify the variations of the Contrastive Learning framework zoo. On the left, Contrastive Multiview Coding (CMC, \citealt{CMC}) stores the representations computed in previous mini-batches in a memory bank to increase the number of negative examples. Although CMC proposes a twin network for comparing distinct versions of the input data, we represent in the figure only one branch to exemplify the memory bank setup. Second from the left, Momentum Contrastive (MoCo, \citealt{MoCo1}) has a parallel network that provides a queue of updated representations of negative examples. In the centre, Simple Contrastive Learning of visual Representations (SimCLR, \citealt{SimCLR1}) uses a unique CNN with a projection head ($h$) to compute the loss and extracts the final representations from the layer before $h$ after training. Second from the right, Boostrap Your Own Latent (BYOL, \citealt{BYOL}) uses a parallel network to compute with the stop gradient operation on one branch to compute a positive pairs-only contrastive loss. On the right, Contrastive Language Image Pre-training (CLIP,~\citealp{2021arXiv210300020R}) trains simultaneously language and visual encoders to generate representations that match both domains.}
    \label{fig:frameworks}
\end{figure*}

\section{Applications of contrastive learning in astronomy}
\label{sec:applications}

Table~\ref{table:applications} summarizes applications of deep CL to astrophysics so far. The first publications are from 2021, which illustrates the fact that the success of CL is relatively recent. Interestingly, the majority of the applications are in the field of galaxy formation. A possible explanation could be that there are large imaging datasets publicly and easily accessible.

Overall, CL is used either for pure data exploration - in which case it is usually followed by some sort of clustering - or to perform a downstream supervised classification or regression / inference using the representations obtained.

\subsection{Inference from representations} 

\subsubsection{Classification with limited amount of labels}
\cite{Hayat-2021} first applied the CL framework MoCo (see Figure~\ref{fig:frameworks} and Subsection~\ref{Sec:frameworks}) to a set of images from the SDSS I and II. They used the learned representations to perform two downstream tasks: galaxy morphology classification and photometric redshift estimation, as well as for data visualization. The work illustrated two main advantages of self-supervised CL. First, the authors showed the importance of using domain knowledge to obtain more robust representations. For example, they introduced custom augmentation to marginalize over reddening. As discussed in Section~\ref{sec:contrastive}, CL easily allows for new augmentations to be included, making it a very flexible framework for representation learning. Second, \cite{Hayat-2021} showed that when the representations are used for galaxy classification and photometric redshift estimation, one can achieve similar accuracies as with a pure supervised approach but with an order of magnitude fewer objects (Figure~\ref{fig:hayat_labels}). This confirms that contrastive learning is helpful for reducing the volume of labeled datasets in astrophysics, as was demonstrated with natural images.

\begin{figure*}
\centering
\includegraphics[width=\textwidth]{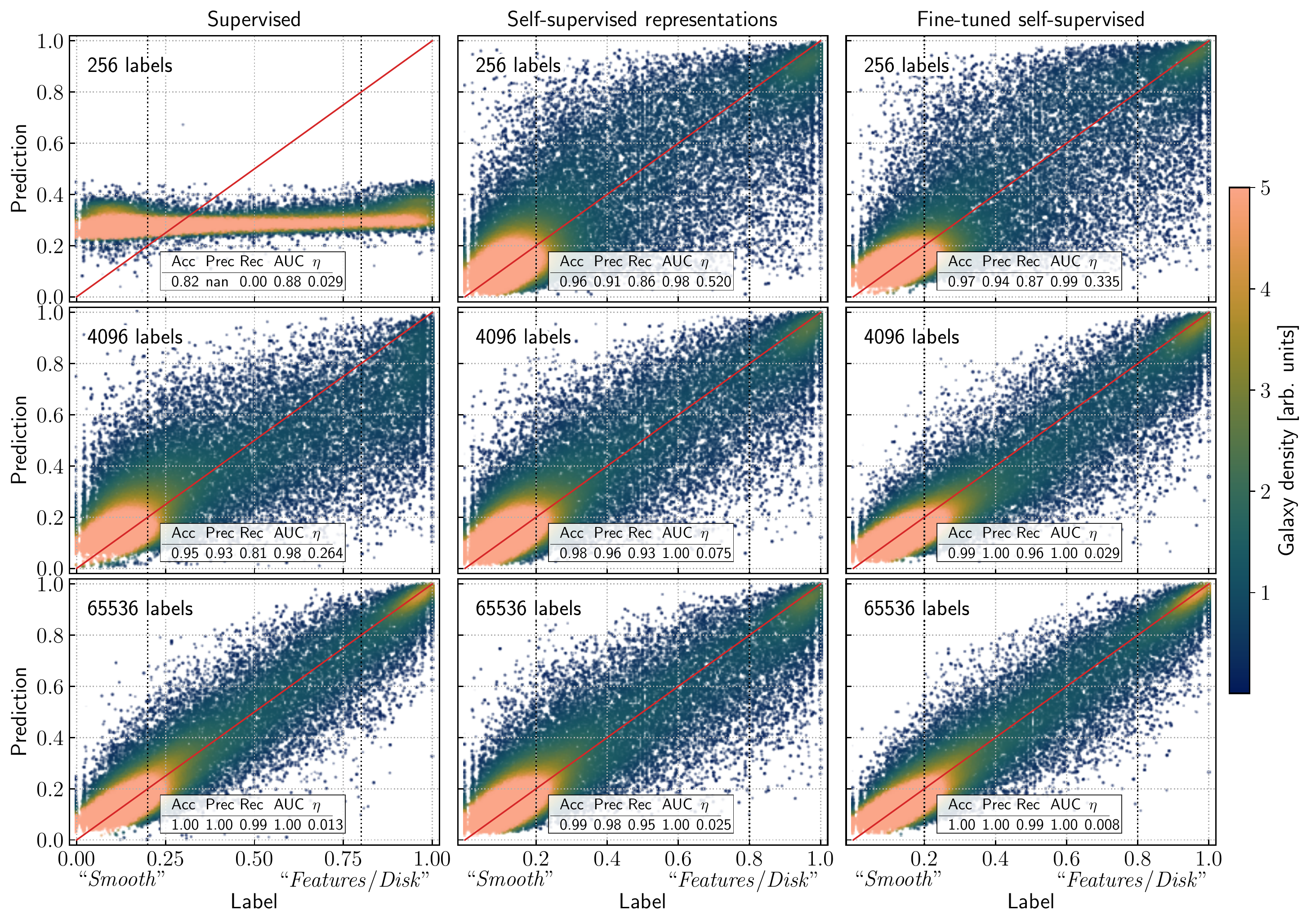}
\caption{Application of contrastive learning (MoCo) to galaxy morphology classification by~\protect\cite{Hayat-2021}. The three columns show the classification performance in a supervised setting (left), a linear classifier directly on the self-supervised representations (center), and when fine-tuning the self-supervised encoder for a few epochs (right). $\eta$ measures the outlier fraction. While a pure supervised setting fails at providing meaningful morphology estimations with 256 examples (top left) a supervised model trained on the representations obtains significantly more accurate results (top middle). }
\label{fig:hayat_labels}
\end{figure*}

A similar conclusion is reached by~\cite{Slijepcevic-2022-Neurips} in an application to radio galaxy classification. By using the BYOL framework (see Figure~\ref{fig:frameworks} and subsection~\ref{Sec:frameworks}) they show that a classification based on the representations achieves comparable accuracy to a complete supervised approach. \cite{Lamdouar-2022} explore the accuracy of CL for classifying time series of solar magnetic field measurements. They show again that CL is an efficient way of obtaining high classification accuracies when only limited labeled data is available (see also~\cite{sunquake_detect} for an extensive analysis of CL representations to classify seismic emissions in the solar surface).

\subsubsection{Domain adaptation}

\cite{Wei-2022} also apply CL followed by a morphological classification downstream task, reaching similar conclusions, i.e., accuracies comparable to supervised classifications are reached but with a small amount of labels. Interestingly, they also show that the representations extracted with self-supervised learning generalize well to multiple imaging datasets. The concept of generalization to multiple tasks (i.e., domain adaptation and generalization) is an important property of CL. Domain adaptation with NNs generally refers to techniques used to improve the performance of a NN model trained on a source domain when applied to a different but related unlabeled target domain, by reducing the discrepancy between the two domains (see \citealp{li2021survey} for a review of the topic). \cite{Walmsley-2022-2} explore this in more detail for galaxy morphology applications. They use, in particular, what they call a \emph{hybrid} contrastive-supervised approach to perform galaxy classifications from a set of visually classified galaxy images. This is performed by adding a supervised term to the BYOL framework, allowing for the performance of classifications while extracting representations that remain invariant to perturbations. They show that the representations learned by the hybrid approach can be efficiently fine-tuned with a few labels to perform supervised tasks for which no labels were provided. They apply this to find new ring galaxy candidates (Figure~\ref{fig:walmsley_hybrid}). In a more recent work,~\cite{2023arXiv230202005C} also explore the concept of a \emph{universal domain adaptation} approach for galaxy morphology. Instead of using a pure CL framework as the ones discussed in this review, they propose instead a custom approach based on an Adaptive Clustering loss~\citep{2020arXiv200207953S,2021arXiv210409415L} on the latent representation. They test their approach by transferring a trained model from SDSS to the Dark Energy Camera Legacy Survey (DECaLS;~\citealp{2019AJ....157..168D}).

\begin{figure}
    \centering
    \includegraphics[width=\hsize]{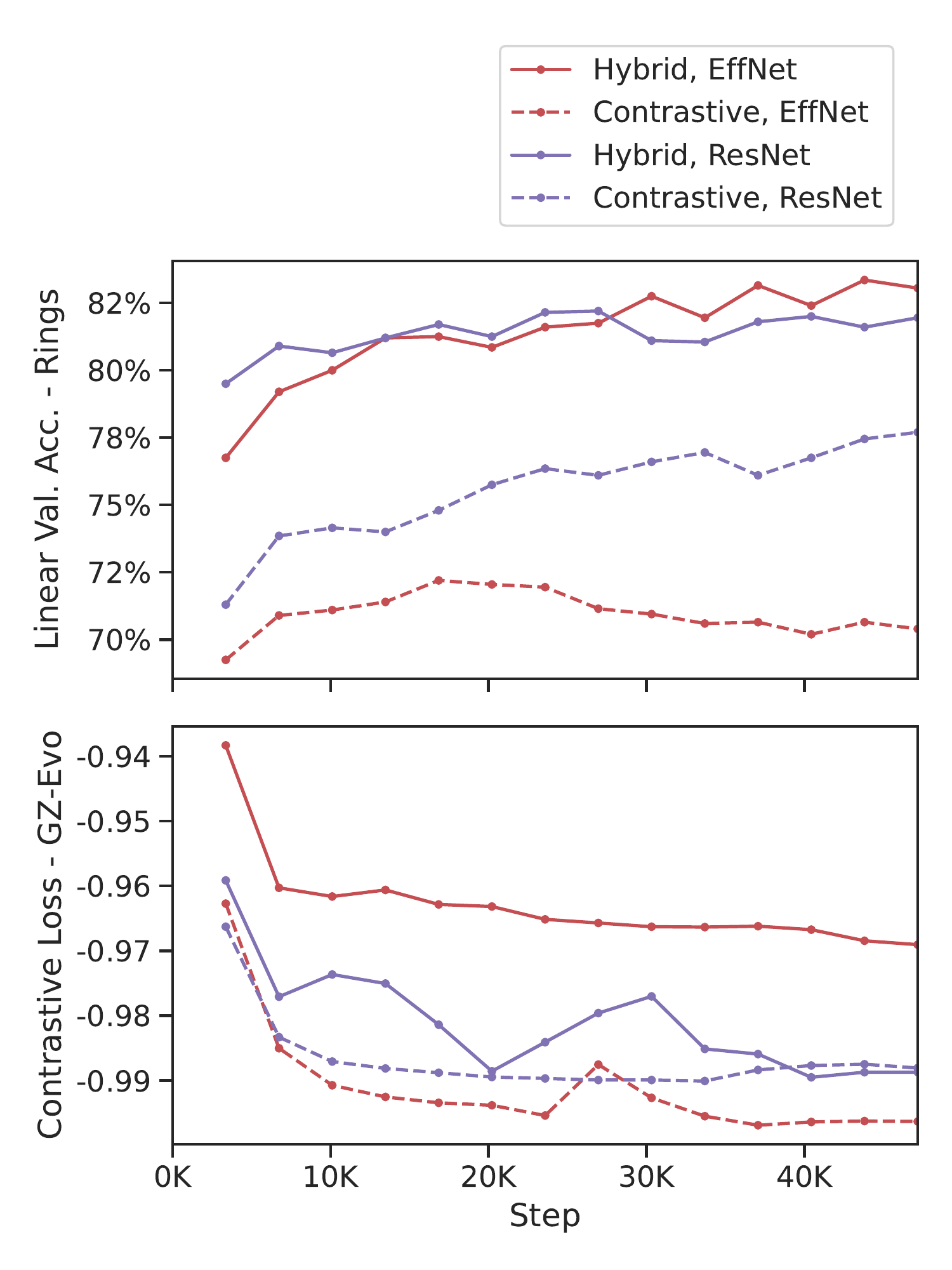}
    \caption{Performance of an hybrid CL approach to identify ring galaxies. The top and bottom panel show the accuracy and the evolution of the contrastive loss function of the downstream classification as a function of training steps. The hybrid CL/supervised approach reaches even higher accuracies than a pure CL one. Figure from~\protect\cite{Walmsley-2022-2}.}
    \label{fig:walmsley_hybrid}
\end{figure}

\subsubsection{Probabilistic inference}
In addition to classification, CL representations can be employed as summary statistics for probabilistic inference. This is the case for example of the work by~\cite{Shen-2022} in which the CL representations are coupled to a Neural Density Estimator (Neural Flow) to estimate an approximate posterior distribution of  physical properties of black hole mergers given the gravitational wave emission. In this case, using CL as conditioning for the Neural Flow has the advantage of improving the inference robustness to noise since the augmentations enable a marginalization over S/N. 


\subsection{Data visualization and clustering}

Another set of applications is oriented towards data visualization and exploration. In these cases, a specific downstream task is not sought, but the main purpose of the representations is to explore and find patterns in the data. In that respect, it is closer to a purely unsupervised application.

\cite{Sarmiento-2021}, for example, use SimCLR (see Figure~\ref{fig:frameworks} and Subsection~\ref{Sec:frameworks}) to extract meaningful representations of inferred stellar population and kinematic maps for $\sim10,000$ galaxies in the Mapping Nearby Galaxies at Apache Point Observatory (MaNGA; \citealp{2015ApJ...798....7B,2022ApJS..259...35A}) survey. They show that the contrastive framework naturally orders galaxies based on their physical properties without supervision and \emph{rediscovers} some well-known relations with a purely data-driven approach. Interestingly, they also find that, for that particular case, other representation learning strategies such as PCA fail to extract physical properties and focus more on instrumental effects. The flexibility of contrastive learning to include custom domain-driven augmentations is key to obtain more physical representations (Figure~\ref{fig:sarmiento_PCA}). More recently,\cite{2023arXiv230207277V} has applied a similar approach to images of galaxies from the Cosmic Evolution Early Release Science survey (CEERS,\citealp{2022ApJ...940L..55F}). By introducing an augmentation that goes from a noiseless to a noisy version of images, they show that the CL representations offer a morphological description that becomes robust to noise, a property lacking in most previous morphology classifications.

\begin{figure*}
    \centering
    \includegraphics[width=0.8\hsize]{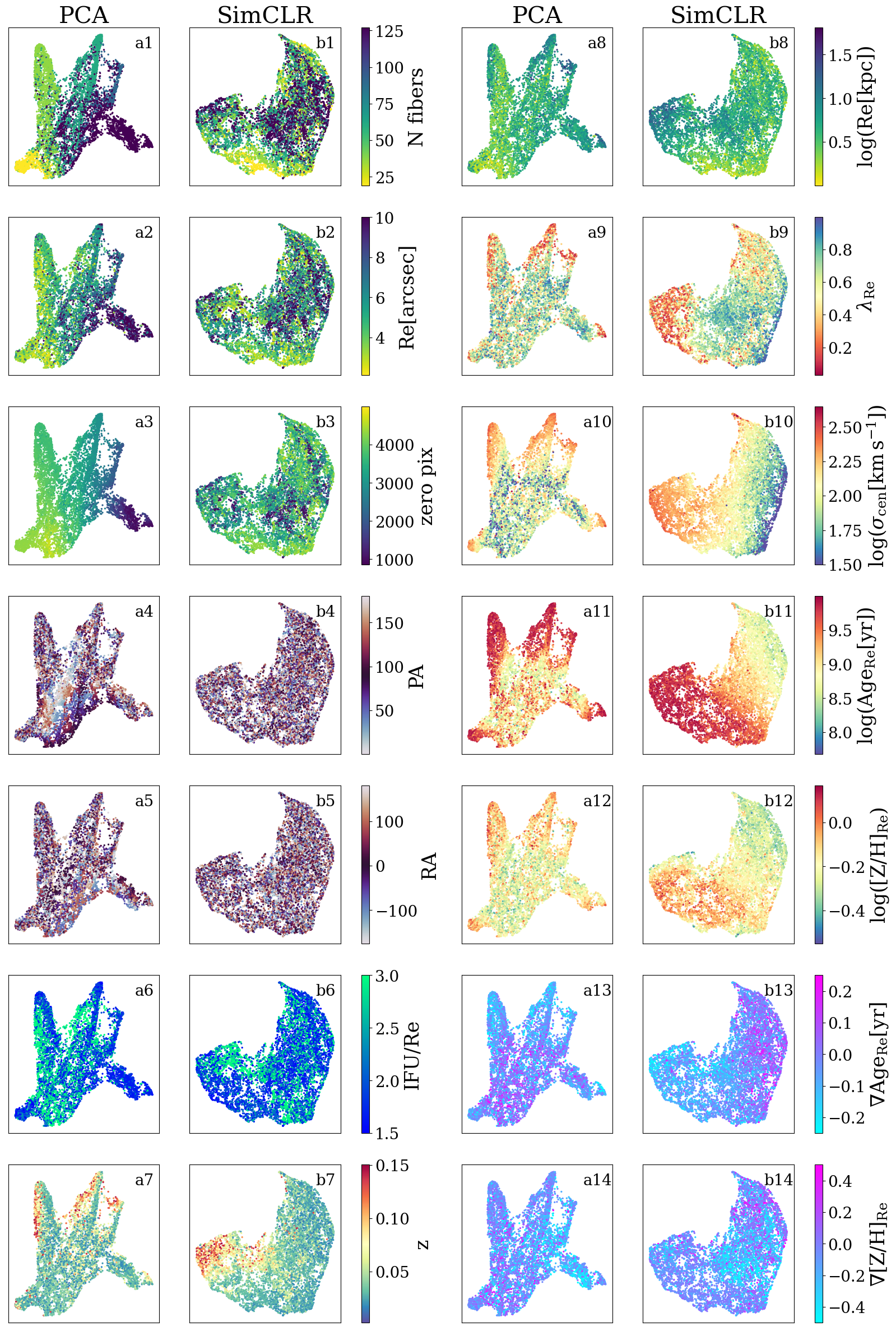}
    \caption{Application of contrastive learning (SimCLR) for data visualization and clustering from~\protect\cite{Sarmiento-2021}. The different panels show a UMAP projection of PCA and SimCLR representations color coded by instrumental and other nuisance parameters (leftmost columns) and physical properties (rightmost columns). While PCA capture mostly instrumental effects, the CL representation are able to marginalize over those.  }
    \label{fig:sarmiento_PCA}
\end{figure*}

\cite{Guo-2022} use a similar contrastive-learning (BYOL) followed by clustering applied to galaxy images in the far-infrared from the Wide-field Infrared Survey Explorer (WISE) survey~\citep{2010AJ....140.1868W}. They demonstrate that the method successfully organizes objects based on similarity and discuss the properties for the different obtained clusters. The work does not, however, discuss the advantages of CL over other representation learning settings in this particular case.

\subsection{Similarity search and anomaly detection}

Two straightforward applications of robust representations are similarity search and anomaly detection. Given that CL representations, by construction, are similar for objects with similar properties - marginalized over the nuisance properties encoded in the augmentations - one can query the representations to look for similar data or for isolated objects. \cite{Stein-2021-Neurips} demonstrate efficient similarity search using CL representations, as illustrated in Figure~\ref{fig:stein_similarity} where the model is queried to identify galaxy images with similar properties.  

\begin{figure*}
    \centering
    \includegraphics[width=\hsize]{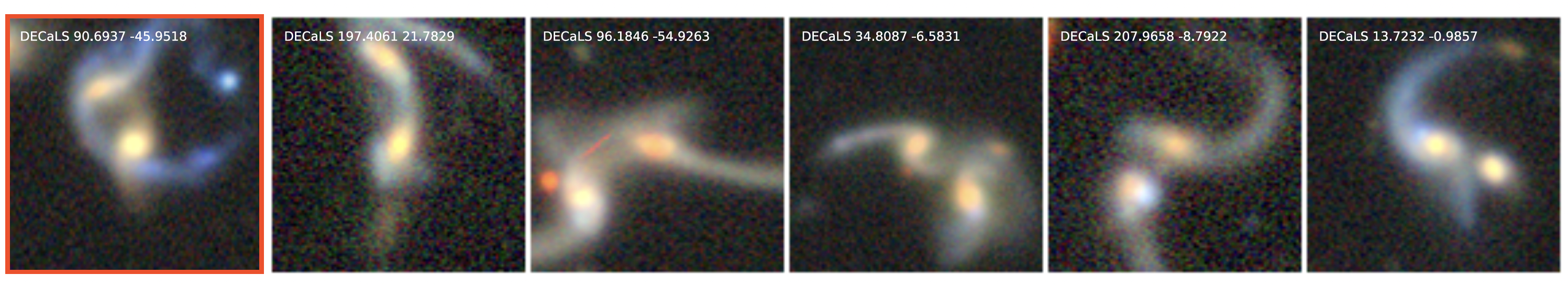}
    \caption{Application of contrastive learning for similarity search of galaxy images with similar morphological features. The leftmost image shows a galaxy candidate and the others are similar image candidates based on the representations. Figure adapted from~\protect\cite{Stein-2021-Neurips}}.
    \label{fig:stein_similarity}
\end{figure*}

\subsection{Other applications}

Recent work by \cite{Doorenbos-2022} uses CL for a different purpose: generating galaxy spectra from images. This illustrates how meaningful representations can be used for a wide number of applications and shows a possible evolution for the near future. The authors use a conditional diffusion model (e.g., \citealp{chen2021conditional}) to generate possible candidates of spectra given an image of a galaxy. Diffusion models are a class of generative models that learn the dynamics of how a distribution evolves over time by iteratively applying a diffusion process to a simple initial distribution, making them well-suited for tasks such as density estimation, generative modeling, and image synthesis (see the introduction section for more references on this type of generative models). Then they use CL to find common representations of real spectra and their corresponding images. Among all candidates resulting from the sampling of the diffusion model, the best candidate is selected by identifying the spectrum producing the closest CL representation to the target image (Figure~\ref{fig:doorenbos_generative}). In addition to the specific application, which is potentially interesting for data exploration, this work illustrates the unique capability of CL to obtain simultaneous representations of heterogeneous data such as spectra and images.

\begin{figure*}
    
    \includegraphics[scale=0.36]{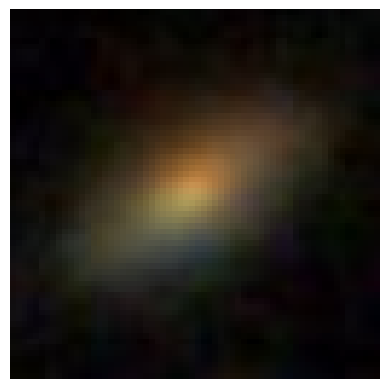}
    \includegraphics[width=0.39\hsize]{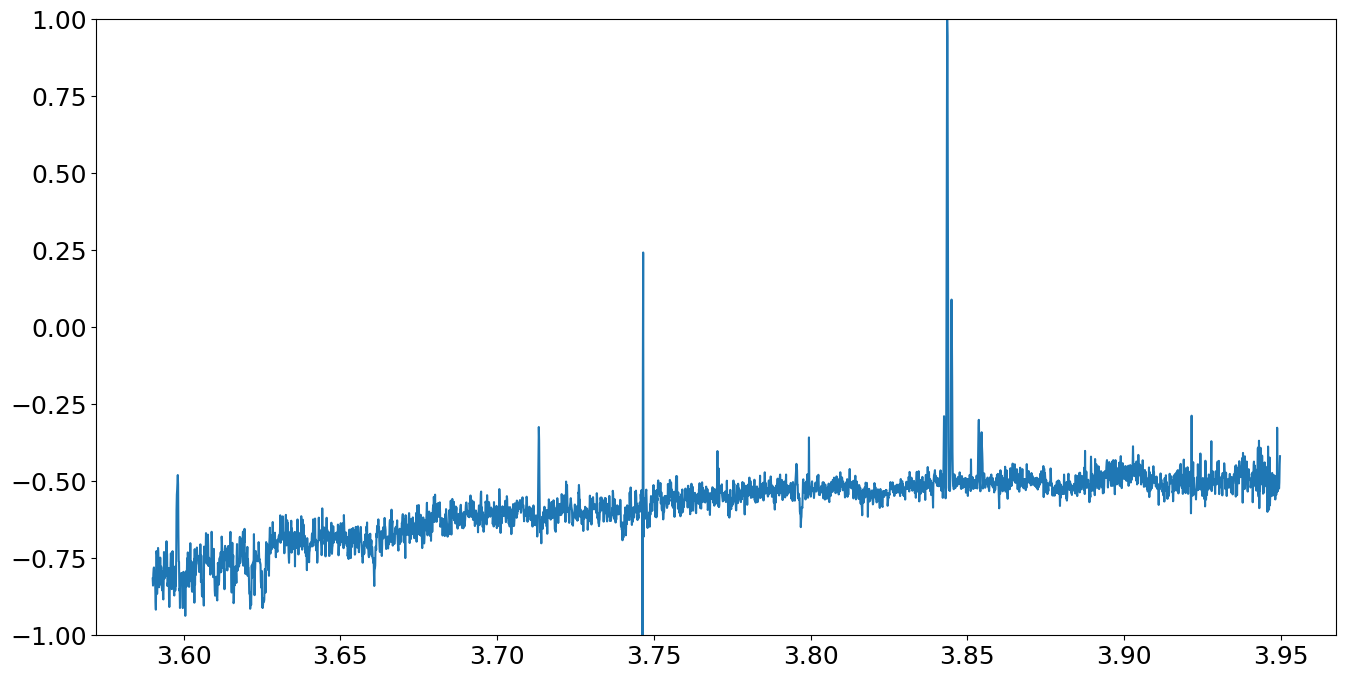}
    \includegraphics[width=0.39\hsize]{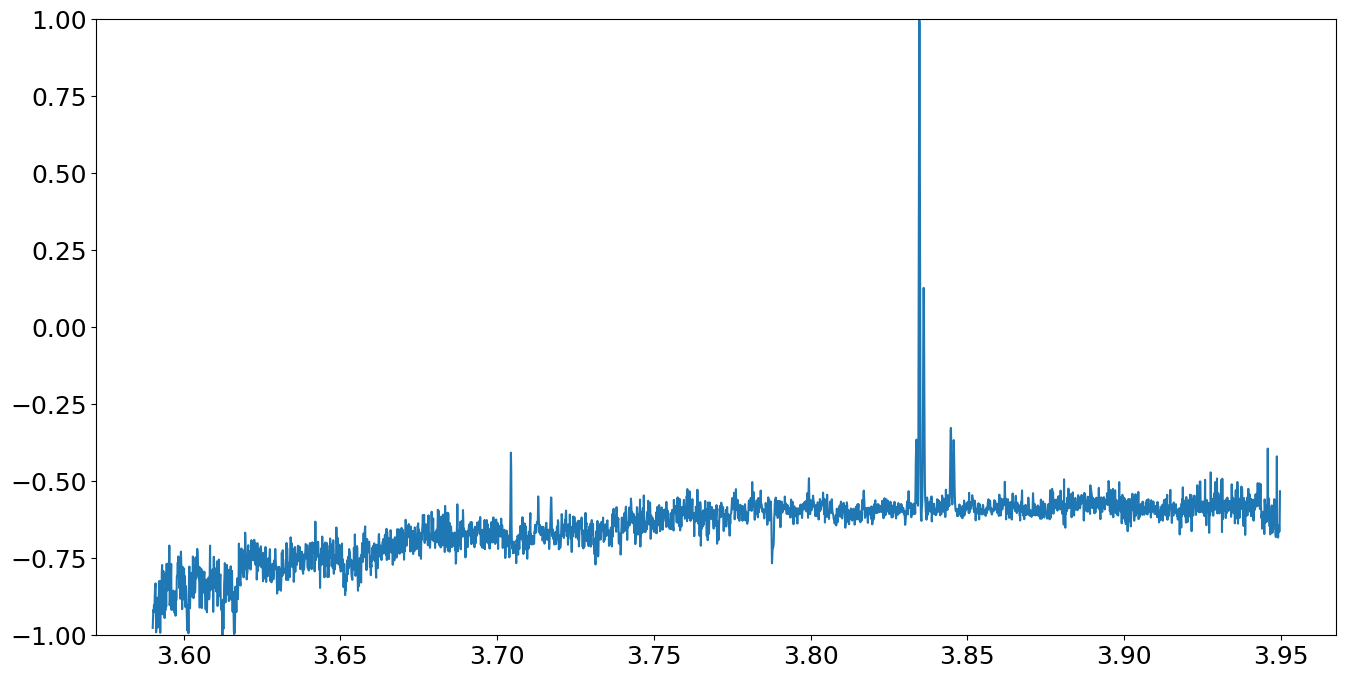}
    \caption{Application of contrastive learning for spectra generation from images. The leftmost column shows the image from SDSS for which a spectrum is predicted. The middle column shows the target spectrum and the rightmost one shows the best candidate according to CL. Figure adapted from~\protect\cite{Doorenbos-2022}. }
    \label{fig:doorenbos_generative}
\end{figure*}

\begin{table*}
\caption{CL applied to astrophysics}             
\label{table:applications}      
\centering                          
\begin{tabular}{l c c c p{4cm}}  
\hline\hline                
Article & Astronomy field & Data & Based on & Downstream task \\    
\hline                      
   \cite{Hayat-2021} & Galaxies & Photometric Imaging & MoCo & Regression (photo-z, morphology) \\  
   &&(SDSS)&\cite{MoCo1}& \\
   \noalign{\smallskip}
   \cite{Sarmiento-2021} & Galaxies & Integral Field Spectroscopy & SimCLR &  Clustering \\
   &&(MaNGA)&\cite{SimCLR1}&\\
   \noalign{\smallskip}
   \cite{Stein-2021-Neurips} & Galaxies & Photometric Imaging & MoCo & Similarity search \\
   &&(DESI Legacy Imaging Survey)&\cite{MoCo1}&\\
   \cite{Walmsley-2022-2} & Galaxies & Photometric Imaging & BYOL (loss) & Transfer Learning\\
   &&&\cite{BYOL}&\\
   \noalign{\smallskip}
   \cite{Slijepcevic-2022-Neurips}&Galaxies&& BYOL & Morphology classification of radio galaxies\\
   &&&\cite{BYOL}&\\
   \noalign{\smallskip}
   \cite{Wei-2022}& Galaxies & Photometric Imaging & SimCLR &Morphology classification\\
   &&SDSS&\cite{SimCLR1}&\\
   \noalign{\smallskip}
   \cite{Shen-2022}& Gravitational Waves & surrogate waveforms &SimCLR (NT-Xent loss)& Regression (masses, spin, quasinormal modes)\\
   &&simulations + LIGO&\cite{SimCLR1}&\\
   \noalign{\smallskip}
   \cite{Doorenbos-2022}& Galaxies & Photometric Imaging and Spectra & SimCLR (NT-Xent loss)& Predict spectra from images (combined with multimodal conditional diffusion models)\\
   &&(SDSS)&\cite{SimCLR1}&\\
   \noalign{\smallskip}
   \cite{Lamdouar-2022}&Solar &&(Triplet loss)&\\
   &&&&\\
   \noalign{\smallskip}
   \cite{Guo-2022} & Galaxies & Photometric Imaging & BYOL &Clustering \\
   &&(WISE)& \cite{BYOL} &\\
   \noalign{\smallskip}
   \cite{sunquake_detect} & Solar & Photospheric Dopplergram maps &SimCLR\& SupCon& Classification (sunquake detection)\\
   && Michelson Doppler Imager &\cite{SimCLR1}&\\
   && Helioseismic and Magnetic Imager & \cite{SupCon-2020} &\\
   \noalign{\smallskip}
   \cite{2023arXiv230207277V} & Galaxies & Photometric Imaging & SimCLR & Domain adaptation and Clustering\\
   &&JWST and TNG50 simulations &\cite{SimCLR1}&\\
   \noalign{\smallskip}
   \cite{2023arXiv230202005C} & Galaxies & Photometric Imaging & Adaptive Clustering & Domain adaptation and Clustering\\
   &&  SDSS and DECaLS & &\\
   \noalign{\smallskip}
   
\hline                                   
\end{tabular}
\end{table*}

\section{Discussion}
\label{sec:discussion}

CL has emerged as an efficient approach for representation learning over the past years with some successful applications in astronomy as described in the previous section. We discuss in the following some practical considerations for non-expert readers interested in using CL methods for other astrophysical problems. 

\subsection{What is contrastive learning useful for?}

CL has  been shown to be primarily a promising way to address the issue of the lack of labeled data for supervised classification and inference.

Over the past few years, deep learning applications have become more frequent in astronomy, with supervised approaches still being the most common (e.g.,\citealp{2023PASA...40....1H}). One of the main bottlenecks of such applications is the availability of labeled data. The vast majority of astronomical data is unlabeled, limiting the applicability of ML for classification or inference. In this context, the community needs methods that can work on small datasets when only a limited amount of labels is available or that can generalize well across domains and, therefore, be applied to different datasets (see e.g.,~\citealp{2023arXiv230202005C}). From the few works exploring CL in astronomy, it appears as a promising approach to at least partially solve some of these issues. For instance,\cite{Hayat-2021} have shown that CL can reduce the number of required labels for supervised classification and regression by an order of magnitude. Along the same lines,\cite{Walmsley-2022-2} have shown that adding a CL component seems to be a promising avenue towards general "foundation" models for morphology classification, i.e., models that can be applied to a variety of different datasets. In addition to being a feature extraction method for supervised downstream tasks, CL can also be used alone as a visualization/exploratory tool (see e.g.,\citealp{Sarmiento-2021,Stein-2021-Neurips}). Although there are various techniques for representation learning (see the introduction) and for transferring trained models (see e.g.,\citealp{2019MNRAS.484...93D,2020A&C....3200390C}), CL has unique properties that make it particularly useful for astrophysics.

 Arguably, the main advantage of CL is that, unlike other approaches, it does not have a predefined encoded metric, allowing it to be adapted for every application using domain knowledge. The similarity pretext task that CL solves is determined by the augmentations, which defines the positive pairs. In that respect, although there are some standard augmentations that can be used for imaging applications (see Figure~\ref{fig:aug-galaxy_img}), CL has room for domain-specific augmentations. This allows one to introduce domain knowledge in the representations without explicitly defining a metric. This is particularly relevant for astrophysics, in which there usually exist selection and instrumental effects that are known and fairly easy to model, but can potentially bias other representation learning algorithms with a fixed metric. CL enables a relatively straightforward marginalization over these nuisance parameters. Several works have already shown the advantages of using domain-specific augmentation. \cite{sunquake_detect} discuss the use of different transformations on regression power maps derived from solar photospheric dopplergrams. They use time-based mixing (stacking of subsequent maps), a low-pass filter followed by solarization, and random erasing of signals.~\cite{Sarmiento-2021} use specific observational effects from the MaNGA survey, while~\cite{Hayat-2021-Neurips} include reddening as part of the augmentation sets. Generally speaking, any dataset affected by known nuisance effects that can be modeled might potentially be a good candidate for application of CL.

Another important consequence of the flexibility of CL for defining the similarity metric is that it opens the door for obtaining representations of different data types. There are a number of recent successful applications of CL in the ML community combining speech and images, for example (e.g.,~\citealp{2021arXiv210212092R}). In astronomy, it is potentially an interesting avenue to extract simultaneous information from images and spectra, a task that is difficult to achieve with other approaches. The first work to have explored this, to our knowledge, is the one by~\cite{Doorenbos-2022}, with encouraging results that open new, interesting research avenues.

\subsection{What is contrastive learning not useful for?}

It is important to emphasize that CL is not particularly efficient as a dimensionality reduction algorithm. In fact, several works have shown that increasing the dimensionality of the representation space generally results in an improved performance (e.g.,\citealp{SimCLR1,Sarmiento-2021}). As a result, the obtained representations are still of fairly high dimension and not easy to visualize. This is why several authors have coupled the resulting representations with a pure dimensionality reduction algorithm for visualization purposes (see e.g.~\citealp{Hayat-2021}). Therefore CL is not the approach to use for inferring the true dimensionality of a dataset. In fact, unlike PCA, there are no constraints on the representations obtained and therefore it is always the case that the different representations are highly correlated.  

This also implies a more challenging interpretability of the extracted features compared to other representation learning algorithms. Both the high dimensionality and the correlation between features significantly complicate the task of physically interpreting the extracted representations. CL is probably not the preferred approach for gaining physical insight by exploring the representations. However, CL frameworks can be coupled with other methods to increase interpretability. One could also potentially add regularization terms into the loss functions to penalize strong correlations in the latent space.

\section{Summary and Conclusions}

This work briefly reviews contrastive methods for representation learning and their application to astrophysics. CL is a self-supervised learning method which obtains meaningful representations of high-dimensionality datasets by solving a similarity pretext task. Compared to other representation learning techniques, the similarity metric is not encoded but is defined through augmentations of the input dataset. By defining \emph{positive pairs} as the set of all augmented versions of the same data point, the representations become invariant to such transformations. 

We review a dozen existing works using CL in astrophysics, the vast majority in the field of galaxy formation. Applications are roughly divided in two main families: downstream regression, inference and classification; and data exploration. From these works, we conclude that CL is as a powerful method to train supervised algorithms with a limited amount of labels. The flexibility of the augmentations allowing domain-specific perturbations makes CL a suitable method for marginalizing over known instrumental or other nuisance effects which can be modeled in astronomy. An interesting avenue for the future, which is also enabled by CL and has just started to be explored in astrophysics, is the combinations of different data types such as imaging and spectra.

\section*{Acknowledgements}

The authors acknowledge financial support from the State Research Agency (AEI\-MCINN) of the Spanish Ministry of Science and Innovation under the grant  "Galaxy Evolution with Artificial Intelligence" with reference PGC2018-100852-A-I00 and under the grant "The structure and evolution of galaxies and their central regions" with reference PID2019-105602GB-I00/10.13039/501100011033, from the ACIISI, Consejer\'{i}a de Econom\'{i}a, Conocimiento y Empleo del Gobierno de Canarias and the European Regional Development Fund (ERDF) under grants with reference PROID2020010057 and PROID2021010044, and from IAC projects P/300724 and P/301802, financed by the Ministry of Science and Innovation, through the State Budget and by the Canary Islands Department of Economy, Knowledge and Employment, through the Regional Budget of the Autonomous Community. 

\section*{Data Availability}

No data are used in this review work.



\bibliographystyle{rasti}
\bibliography{biblio} 

\begin{thebibliography}{85}
\expandafter\ifx\csname natexlab\endcsname\relax\def\natexlab#1{#1}\fi

\bibitem[{Abazajian} et~al.(2009){Abazajian}, {Adelman-McCarthy},
  {Ag{\"u}eros}, {Allam}, {Allende Prieto}, {An}, {Anderson}, {Anderson},
  {Annis}, {Bahcall}, {Bailer-Jones}, {Barentine}, {Bassett}, {Becker},
  {Beers}, {Bell}, {Belokurov}, {Berlind}, {Berman}, {Bernardi}, {Bickerton},
  {Bizyaev}, {Blakeslee}, {Blanton}, {Bochanski}, {Boroski}, {Brewington},
  {Brinchmann}, {Brinkmann}, {Brunner}, {Budav{\'a}ri}, {Carey}, {Carliles},
  {Carr}, {Castander}, {Cinabro}, {Connolly}, {Csabai}, {Cunha}, {Czarapata},
  {Davenport}, {de Haas}, {Dilday}, {Doi}, {Eisenstein}, {Evans}, {Evans},
  {Fan}, {Friedman}, {Frieman}, {Fukugita}, {G{\"a}nsicke}, {Gates},
  {Gillespie}, {Gilmore}, {Gonzalez}, {Gonzalez}, {Grebel}, {Gunn},
  {Gy{\"o}ry}, {Hall}, {Harding}, {Harris}, {Harvanek}, {Hawley}, {Hayes},
  {Heckman}, {Hendry}, {Hennessy}, {Hindsley}, {Hoblitt}, {Hogan}, {Hogg},
  {Holtzman}, {Hyde}, {Ichikawa}, {Ichikawa}, {Im}, {Ivezi{\'c}}, {Jester},
  {Jiang}, {Johnson}, {Jorgensen}, {Juri{\'c}}, {Kent}, {Kessler}, {Kleinman},
  {Knapp}, {Konishi}, {Kron}, {Krzesinski}, {Kuropatkin}, {Lampeitl},
  {Lebedeva}, {Lee}, {Lee}, {French Leger}, {L{\'e}pine}, {Li}, {Lima}, {Lin},
  {Long}, {Loomis}, {Loveday}, {Lupton}, {Magnier}, {Malanushenko},
  {Malanushenko}, {Mandelbaum}, {Margon}, {Marriner}, {Mart{\'\i}nez-Delgado},
  {Matsubara}, {McGehee}, {McKay}, {Meiksin}, {Morrison}, {Mullally}, {Munn},
  {Murphy}, {Nash}, {Nebot}, {Neilsen}, {Newberg}, {Newman}, {Nichol},
  {Nicinski}, {Nieto-Santisteban}, {Nitta}, {Okamura}, {Oravetz}, {Ostriker},
  {Owen}, {Padmanabhan}, {Pan}, {Park}, {Pauls}, {Peoples}, {Percival}, {Pier},
  {Pope}, {Pourbaix}, {Price}, {Purger}, {Quinn}, {Raddick}, {Re Fiorentin},
  {Richards}, {Richmond}, {Riess}, {Rix}, {Rockosi}, {Sako}, {Schlegel},
  {Schneider}, {Scholz}, {Schreiber}, {Schwope}, {Seljak}, {Sesar}, {Sheldon},
  {Shimasaku}, {Sibley}, {Simmons}, {Sivarani}, {Allyn Smith}, {Smith},
  {Smol{\v{c}}i{\'c}}, {Snedden}, {Stebbins}, {Steinmetz}, {Stoughton},
  {Strauss}, {SubbaRao}, {Suto}, {Szalay}, {Szapudi}, {Szkody}, {Tanaka},
  {Tegmark}, {Teodoro}, {Thakar}, {Tremonti}, {Tucker}, {Uomoto}, {Vanden
  Berk}, {Vandenberg}, {Vidrih}, {Vogeley}, {Voges}, {Vogt}, {Wadadekar},
  {Watters}, {Weinberg}, {West}, {White}, {Wilhite}, {Wonders}, {Yanny},
  {Yocum}, {York}, {Zehavi}, {Zibetti}, \& {Zucker}]{2009ApJS..182..543A}
{Abazajian}, K.~N., {Adelman-McCarthy}, J.~K., {Ag{\"u}eros}, M.~A., {Allam},
  S.~S., {Allende Prieto}, C., {An}, D., {Anderson}, K. S.~J., {Anderson},
  S.~F., {Annis}, J., {Bahcall}, N.~A., {Bailer-Jones}, C.~A.~L., {Barentine},
  J.~C., {Bassett}, B.~A., {Becker}, A.~C., {Beers}, T.~C., {Bell}, E.~F.,
  {Belokurov}, V., {Berlind}, A.~A., {Berman}, E.~F., {Bernardi}, M.,
  {Bickerton}, S.~J., {Bizyaev}, D., {Blakeslee}, J.~P., {Blanton}, M.~R.,
  {Bochanski}, J.~J., {Boroski}, W.~N., {Brewington}, H.~J., {Brinchmann}, J.,
  {Brinkmann}, J., {Brunner}, R.~J., {Budav{\'a}ri}, T., {Carey}, L.~N.,
  {Carliles}, S., {Carr}, M.~A., {Castander}, F.~J., {Cinabro}, D., {Connolly},
  A.~J., {Csabai}, I., {Cunha}, C.~E., {Czarapata}, P.~C., {Davenport}, J.
  R.~A., {de Haas}, E., {Dilday}, B., {Doi}, M., {Eisenstein}, D.~J., {Evans},
  M.~L., {Evans}, N.~W., {Fan}, X., {Friedman}, S.~D., {Frieman}, J.~A.,
  {Fukugita}, M., {G{\"a}nsicke}, B.~T., {Gates}, E., {Gillespie}, B.,
  {Gilmore}, G., {Gonzalez}, B., {Gonzalez}, C.~F., {Grebel}, E.~K., {Gunn},
  J.~E., {Gy{\"o}ry}, Z., {Hall}, P.~B., {Harding}, P., {Harris}, F.~H.,
  {Harvanek}, M., {Hawley}, S.~L., {Hayes}, J. J.~E., {Heckman}, T.~M.,
  {Hendry}, J.~S., {Hennessy}, G.~S., {Hindsley}, R.~B., {Hoblitt}, J.,
  {Hogan}, C.~J., {Hogg}, D.~W., {Holtzman}, J.~A., {Hyde}, J.~B., {Ichikawa},
  S.-i., {Ichikawa}, T., {Im}, M., {Ivezi{\'c}}, {\v{Z}}., {Jester}, S.,
  {Jiang}, L., {Johnson}, J.~A., {Jorgensen}, A.~M., {Juri{\'c}}, M., {Kent},
  S.~M., {Kessler}, R., {Kleinman}, S.~J., {Knapp}, G.~R., {Konishi}, K.,
  {Kron}, R.~G., {Krzesinski}, J., {Kuropatkin}, N., {Lampeitl}, H.,
  {Lebedeva}, S., {Lee}, M.~G., {Lee}, Y.~S., {French Leger}, R., {L{\'e}pine},
  S., {Li}, N., {Lima}, M., {Lin}, H., {Long}, D.~C., {Loomis}, C.~P.,
  {Loveday}, J., {Lupton}, R.~H., {Magnier}, E., {Malanushenko}, O.,
  {Malanushenko}, V., {Mandelbaum}, R., {Margon}, B., {Marriner}, J.~P.,
  {Mart{\'\i}nez-Delgado}, D., {Matsubara}, T., {McGehee}, P.~M., {McKay},
  T.~A., {Meiksin}, A., {Morrison}, H.~L., {Mullally}, F., {Munn}, J.~A.,
  {Murphy}, T., {Nash}, T., {Nebot}, A., {Neilsen}, Eric~H., J., {Newberg},
  H.~J., {Newman}, P.~R., {Nichol}, R.~C., {Nicinski}, T., {Nieto-Santisteban},
  M., {Nitta}, A., {Okamura}, S., {Oravetz}, D.~J., {Ostriker}, J.~P., {Owen},
  R., {Padmanabhan}, N., {Pan}, K., {Park}, C., {Pauls}, G., {Peoples}, John,
  J., {Percival}, W.~J., {Pier}, J.~R., {Pope}, A.~C., {Pourbaix}, D., {Price},
  P.~A., {Purger}, N., {Quinn}, T., {Raddick}, M.~J., {Re Fiorentin}, P.,
  {Richards}, G.~T., {Richmond}, M.~W., {Riess}, A.~G., {Rix}, H.-W.,
  {Rockosi}, C.~M., {Sako}, M., {Schlegel}, D.~J., {Schneider}, D.~P.,
  {Scholz}, R.-D., {Schreiber}, M.~R., {Schwope}, A.~D., {Seljak}, U., {Sesar},
  B., {Sheldon}, E., {Shimasaku}, K., {Sibley}, V.~C., {Simmons}, A.~E.,
  {Sivarani}, T., {Allyn Smith}, J., {Smith}, M.~C., {Smol{\v{c}}i{\'c}}, V.,
  {Snedden}, S.~A., {Stebbins}, A., {Steinmetz}, M., {Stoughton}, C.,
  {Strauss}, M.~A., {SubbaRao}, M., {Suto}, Y., {Szalay}, A.~S., {Szapudi}, I.,
  {Szkody}, P., {Tanaka}, M., {Tegmark}, M., {Teodoro}, L. F.~A., {Thakar},
  A.~R., {Tremonti}, C.~A., {Tucker}, D.~L., {Uomoto}, A., {Vanden Berk},
  D.~E., {Vandenberg}, J., {Vidrih}, S., {Vogeley}, M.~S., {Voges}, W., {Vogt},
  N.~P., {Wadadekar}, Y., {Watters}, S., {Weinberg}, D.~H., {West}, A.~A.,
  {White}, S. D.~M., {Wilhite}, B.~C., {Wonders}, A.~C., {Yanny}, B., {Yocum},
  D.~R., {York}, D.~G., {Zehavi}, I., {Zibetti}, S., \& {Zucker}, D.~B., 2009.
\newblock {The Seventh Data Release of the Sloan Digital Sky Survey}, {\it
  \apjs\/}, {\bf 182}(2), 543--558.

\bibitem[{Abdurro'uf} et~al.(2022){Abdurro'uf}, {Accetta}, {Aerts}, {Silva
  Aguirre}, {Ahumada}, {Ajgaonkar}, {Filiz Ak}, {Alam}, {Allende Prieto},
  {Almeida}, {Anders}, {Anderson}, {Andrews}, {Anguiano}, {Aquino-Ort{\'\i}z},
  {Arag{\'o}n-Salamanca}, {Argudo-Fern{\'a}ndez}, {Ata}, {Aubert},
  {Avila-Reese}, {Badenes}, {Barb{\'a}}, {Barger}, {Barrera-Ballesteros},
  {Beaton}, {Beers}, {Belfiore}, {Bender}, {Bernardi}, {Bershady}, {Beutler},
  {Bidin}, {Bird}, {Bizyaev}, {Blanc}, {Blanton}, {Boardman}, {Bolton},
  {Boquien}, {Borissova}, {Bovy}, {Brandt}, {Brown}, {Brownstein}, {Brusa},
  {Buchner}, {Bundy}, {Burchett}, {Bureau}, {Burgasser}, {Cabang}, {Campbell},
  {Cappellari}, {Carlberg}, {Wanderley}, {Carrera}, {Cash}, {Chen}, {Chen},
  {Cherinka}, {Chiappini}, {Choi}, {Chojnowski}, {Chung}, {Clerc}, {Cohen},
  {Comerford}, {Comparat}, {da Costa}, {Covey}, {Crane}, {Cruz-Gonzalez},
  {Culhane}, {Cunha}, {Dai}, {Damke}, {Darling}, {Davidson}, {Davies},
  {Dawson}, {De Lee}, {Diamond-Stanic}, {Cano-D{\'\i}az}, {S{\'a}nchez},
  {Donor}, {Duckworth}, {Dwelly}, {Eisenstein}, {Elsworth}, {Emsellem},
  {Eracleous}, {Escoffier}, {Fan}, {Farr}, {Feng}, {Fern{\'a}ndez-Trincado},
  {Feuillet}, {Filipp}, {Fillingham}, {Frinchaboy}, {Fromenteau}, {Galbany},
  {Garc{\'\i}a}, {Garc{\'\i}a-Hern{\'a}ndez}, {Ge}, {Geisler}, {Gelfand},
  {G{\'e}ron}, {Gibson}, {Goddy}, {Godoy-Rivera}, {Grabowski}, {Green},
  {Greener}, {Grier}, {Griffith}, {Guo}, {Guy}, {Hadjara}, {Harding},
  {Hasselquist}, {Hayes}, {Hearty}, {Hern{\'a}ndez}, {Hill}, {Hogg},
  {Holtzman}, {Horta}, {Hsieh}, {Hsu}, {Hsu}, {Huber}, {Huertas-Company},
  {Hutchinson}, {Hwang}, {Ibarra-Medel}, {Chitham}, {Ilha}, {Imig}, {Jaekle},
  {Jayasinghe}, {Ji}, {Johnson}, {Jones}, {J{\"o}nsson}, {Katkov}, {Khalatyan},
  {Kinemuchi}, {Kisku}, {Knapen}, {Kneib}, {Kollmeier}, {Kong}, {Kounkel},
  {Kreckel}, {Krishnarao}, {Lacerna}, {Lane}, {Langgin}, {Lavender}, {Law},
  {Lazarz}, {Leung}, {Leung}, {Lewis}, {Li}, {Li}, {Lian}, {Liang}, {Lin},
  {Lin}, {Lin}, {Lintott}, {Long}, {Longa-Pe{\~n}a}, {L{\'o}pez-Cob{\'a}},
  {Lu}, {Lundgren}, {Luo}, {Mackereth}, {de la Macorra}, {Mahadevan},
  {Majewski}, {Manchado}, {Mandeville}, {Maraston}, {Margalef-Bentabol},
  {Masseron}, {Masters}, {Mathur}, {McDermid}, {Mckay}, {Merloni},
  {Merrifield}, {Meszaros}, {Miglio}, {Di Mille}, {Minniti}, {Minsley},
  {Monachesi}, {Moon}, {Mosser}, {Mulchaey}, {Muna}, {Mu{\~n}oz}, {Myers},
  {Myers}, {Nadathur}, {Nair}, {Nandra}, {Neumann}, {Newman}, {Nidever},
  {Nikakhtar}, {Nitschelm}, {O'Connell}, {Garma-Oehmichen}, {Luan Souza de
  Oliveira}, {Olney}, {Oravetz}, {Ortigoza-Urdaneta}, {Osorio}, {Otter},
  {Pace}, {Padilla}, {Pan}, {Pan}, {Parikh}, {Parker}, {Peirani}, {Pe{\~n}a
  Ram{\'\i}rez}, {Penny}, {Percival}, {Perez-Fournon}, {Pinsonneault},
  {Poidevin}, {Poovelil}, {Price-Whelan}, {B{\'a}rbara de Andrade Queiroz},
  {Raddick}, {Ray}, {Rembold}, {Riddle}, {Riffel}, {Riffel}, {Rix}, {Robin},
  {Rodr{\'\i}guez-Puebla}, {Roman-Lopes}, {Rom{\'a}n-Z{\'u}{\~n}iga}, {Rose},
  {Ross}, {Rossi}, {Rubin}, {Salvato}, {S{\'a}nchez}, {S{\'a}nchez-Gallego},
  {Sanderson}, {Santana Rojas}, {Sarceno}, {Sarmiento}, {Sayres}, {Sazonova},
  {Schaefer}, {Schiavon}, {Schlegel}, {Schneider}, {Schultheis}, {Schwope},
  {Serenelli}, {Serna}, {Shao}, {Shapiro}, {Sharma}, {Shen}, {Shetrone}, {Shu},
  {Simon}, {Skrutskie}, {Smethurst}, {Smith}, {Sobeck}, {Spoo}, {Sprague},
  {Stark}, {Stassun}, {Steinmetz}, {Stello}, {Stone-Martinez},
  {Storchi-Bergmann}, {Stringfellow}, {Stutz}, {Su}, {Taghizadeh-Popp},
  {Talbot}, {Tayar}, {Telles}, {Teske}, {Thakar}, {Theissen}, {Tkachenko},
  {Thomas}, {Tojeiro}, {Hernandez Toledo}, {Troup}, {Trump}, {Trussler},
  {Turner}, {Tuttle}, {Unda-Sanzana}, {V{\'a}zquez-Mata}, {Valentini},
  {Valenzuela}, {Vargas-Gonz{\'a}lez}, {Vargas-Maga{\~n}a}, {Alfaro},
  {Villanova}, {Vincenzo}, {Wake}, {Warfield}, {Washington}, {Weaver},
  {Weijmans}, {Weinberg}, {Weiss}, {Westfall}, {Wild}, {Wilde}, {Wilson},
  {Wilson}, {Wilson}, {Wolf}, {Wood-Vasey}, {Yan}, {Zamora}, {Zasowski},
  {Zhang}, {Zhao}, {Zheng}, {Zheng}, \& {Zhu}]{2022ApJS..259...35A}
{Abdurro'uf}, {Accetta}, K., {Aerts}, C., {Silva Aguirre}, V., {Ahumada}, R.,
  {Ajgaonkar}, N., {Filiz Ak}, N., {Alam}, S., {Allende Prieto}, C., {Almeida},
  A., {Anders}, F., {Anderson}, S.~F., {Andrews}, B.~H., {Anguiano}, B.,
  {Aquino-Ort{\'\i}z}, E., {Arag{\'o}n-Salamanca}, A., {Argudo-Fern{\'a}ndez},
  M., {Ata}, M., {Aubert}, M., {Avila-Reese}, V., {Badenes}, C., {Barb{\'a}},
  R.~H., {Barger}, K., {Barrera-Ballesteros}, J.~K., {Beaton}, R.~L., {Beers},
  T.~C., {Belfiore}, F., {Bender}, C.~F., {Bernardi}, M., {Bershady}, M.~A.,
  {Beutler}, F., {Bidin}, C.~M., {Bird}, J.~C., {Bizyaev}, D., {Blanc}, G.~A.,
  {Blanton}, M.~R., {Boardman}, N.~F., {Bolton}, A.~S., {Boquien}, M.,
  {Borissova}, J., {Bovy}, J., {Brandt}, W.~N., {Brown}, J., {Brownstein},
  J.~R., {Brusa}, M., {Buchner}, J., {Bundy}, K., {Burchett}, J.~N., {Bureau},
  M., {Burgasser}, A., {Cabang}, T.~K., {Campbell}, S., {Cappellari}, M.,
  {Carlberg}, J.~K., {Wanderley}, F.~C., {Carrera}, R., {Cash}, J., {Chen},
  Y.-P., {Chen}, W.-H., {Cherinka}, B., {Chiappini}, C., {Choi}, P.~D.,
  {Chojnowski}, S.~D., {Chung}, H., {Clerc}, N., {Cohen}, R.~E., {Comerford},
  J.~M., {Comparat}, J., {da Costa}, L., {Covey}, K., {Crane}, J.~D.,
  {Cruz-Gonzalez}, I., {Culhane}, C., {Cunha}, K., {Dai}, Y.~S., {Damke}, G.,
  {Darling}, J., {Davidson}, James~W., J., {Davies}, R., {Dawson}, K., {De
  Lee}, N., {Diamond-Stanic}, A.~M., {Cano-D{\'\i}az}, M., {S{\'a}nchez},
  H.~D., {Donor}, J., {Duckworth}, C., {Dwelly}, T., {Eisenstein}, D.~J.,
  {Elsworth}, Y.~P., {Emsellem}, E., {Eracleous}, M., {Escoffier}, S., {Fan},
  X., {Farr}, E., {Feng}, S., {Fern{\'a}ndez-Trincado}, J.~G., {Feuillet}, D.,
  {Filipp}, A., {Fillingham}, S.~P., {Frinchaboy}, P.~M., {Fromenteau}, S.,
  {Galbany}, L., {Garc{\'\i}a}, R.~A., {Garc{\'\i}a-Hern{\'a}ndez}, D.~A.,
  {Ge}, J., {Geisler}, D., {Gelfand}, J., {G{\'e}ron}, T., {Gibson}, B.~J.,
  {Goddy}, J., {Godoy-Rivera}, D., {Grabowski}, K., {Green}, P.~J., {Greener},
  M., {Grier}, C.~J., {Griffith}, E., {Guo}, H., {Guy}, J., {Hadjara}, M.,
  {Harding}, P., {Hasselquist}, S., {Hayes}, C.~R., {Hearty}, F.,
  {Hern{\'a}ndez}, J., {Hill}, L., {Hogg}, D.~W., {Holtzman}, J.~A., {Horta},
  D., {Hsieh}, B.-C., {Hsu}, C.-H., {Hsu}, Y.-H., {Huber}, D.,
  {Huertas-Company}, M., {Hutchinson}, B., {Hwang}, H.~S., {Ibarra-Medel},
  H.~J., {Chitham}, J.~I., {Ilha}, G.~S., {Imig}, J., {Jaekle}, W.,
  {Jayasinghe}, T., {Ji}, X., {Johnson}, J.~A., {Jones}, A., {J{\"o}nsson}, H.,
  {Katkov}, I., {Khalatyan}, Arman, D., {Kinemuchi}, K., {Kisku}, S., {Knapen},
  J.~H., {Kneib}, J.-P., {Kollmeier}, J.~A., {Kong}, M., {Kounkel}, M.,
  {Kreckel}, K., {Krishnarao}, D., {Lacerna}, I., {Lane}, R.~R., {Langgin}, R.,
  {Lavender}, R., {Law}, D.~R., {Lazarz}, D., {Leung}, H.~W., {Leung}, H.-H.,
  {Lewis}, H.~M., {Li}, C., {Li}, R., {Lian}, J., {Liang}, F.-H., {Lin}, L.,
  {Lin}, Y.-T., {Lin}, S., {Lintott}, C., {Long}, D., {Longa-Pe{\~n}a}, P.,
  {L{\'o}pez-Cob{\'a}}, C., {Lu}, S., {Lundgren}, B.~F., {Luo}, Y.,
  {Mackereth}, J.~T., {de la Macorra}, A., {Mahadevan}, S., {Majewski}, S.~R.,
  {Manchado}, A., {Mandeville}, T., {Maraston}, C., {Margalef-Bentabol}, B.,
  {Masseron}, T., {Masters}, K.~L., {Mathur}, S., {McDermid}, R.~M., {Mckay},
  M., {Merloni}, A., {Merrifield}, M., {Meszaros}, S., {Miglio}, A., {Di
  Mille}, F., {Minniti}, D., {Minsley}, R., {Monachesi}, A., {Moon}, J.,
  {Mosser}, B., {Mulchaey}, J., {Muna}, D., {Mu{\~n}oz}, R.~R., {Myers}, A.~D.,
  {Myers}, N., {Nadathur}, S., {Nair}, P., {Nandra}, K., {Neumann}, J.,
  {Newman}, J.~A., {Nidever}, D.~L., {Nikakhtar}, F., {Nitschelm}, C.,
  {O'Connell}, J.~E., {Garma-Oehmichen}, L., {Luan Souza de Oliveira}, G.,
  {Olney}, R., {Oravetz}, D., {Ortigoza-Urdaneta}, M., {Osorio}, Y., {Otter},
  J., {Pace}, Z.~J., {Padilla}, N., {Pan}, K., {Pan}, H.-A., {Parikh}, T.,
  {Parker}, J., {Peirani}, S., {Pe{\~n}a Ram{\'\i}rez}, K., {Penny}, S.,
  {Percival}, W.~J., {Perez-Fournon}, I., {Pinsonneault}, M., {Poidevin}, F.,
  {Poovelil}, V.~J., {Price-Whelan}, A.~M., {B{\'a}rbara de Andrade Queiroz},
  A., {Raddick}, M.~J., {Ray}, A., {Rembold}, S.~B., {Riddle}, N., {Riffel},
  R.~A., {Riffel}, R., {Rix}, H.-W., {Robin}, A.~C., {Rodr{\'\i}guez-Puebla},
  A., {Roman-Lopes}, A., {Rom{\'a}n-Z{\'u}{\~n}iga}, C., {Rose}, B., {Ross},
  A.~J., {Rossi}, G., {Rubin}, K. H.~R., {Salvato}, M., {S{\'a}nchez}, S.~F.,
  {S{\'a}nchez-Gallego}, J.~R., {Sanderson}, R., {Santana Rojas}, F.~A.,
  {Sarceno}, E., {Sarmiento}, R., {Sayres}, C., {Sazonova}, E., {Schaefer},
  A.~L., {Schiavon}, R., {Schlegel}, D.~J., {Schneider}, D.~P., {Schultheis},
  M., {Schwope}, A., {Serenelli}, A., {Serna}, J., {Shao}, Z., {Shapiro}, G.,
  {Sharma}, A., {Shen}, Y., {Shetrone}, M., {Shu}, Y., {Simon}, J.~D.,
  {Skrutskie}, M.~F., {Smethurst}, R., {Smith}, V., {Sobeck}, J., {Spoo}, T.,
  {Sprague}, D., {Stark}, D.~V., {Stassun}, K.~G., {Steinmetz}, M., {Stello},
  D., {Stone-Martinez}, A., {Storchi-Bergmann}, T., {Stringfellow}, G.~S.,
  {Stutz}, A., {Su}, Y.-C., {Taghizadeh-Popp}, M., {Talbot}, M.~S., {Tayar},
  J., {Telles}, E., {Teske}, J., {Thakar}, A., {Theissen}, C., {Tkachenko}, A.,
  {Thomas}, D., {Tojeiro}, R., {Hernandez Toledo}, H., {Troup}, N.~W., {Trump},
  J.~R., {Trussler}, J., {Turner}, J., {Tuttle}, S., {Unda-Sanzana}, E.,
  {V{\'a}zquez-Mata}, J.~A., {Valentini}, M., {Valenzuela}, O.,
  {Vargas-Gonz{\'a}lez}, J., {Vargas-Maga{\~n}a}, M., {Alfaro}, P.~V.,
  {Villanova}, S., {Vincenzo}, F., {Wake}, D., {Warfield}, J.~T., {Washington},
  J.~D., {Weaver}, B.~A., {Weijmans}, A.-M., {Weinberg}, D.~H., {Weiss}, A.,
  {Westfall}, K.~B., {Wild}, V., {Wilde}, M.~C., {Wilson}, J.~C., {Wilson},
  R.~F., {Wilson}, M., {Wolf}, J., {Wood-Vasey}, W.~M., {Yan}, R., {Zamora},
  O., {Zasowski}, G., {Zhang}, K., {Zhao}, C., {Zheng}, Z., {Zheng}, Z., \&
  {Zhu}, K., 2022.
\newblock {The Seventeenth Data Release of the Sloan Digital Sky Surveys:
  Complete Release of MaNGA, MaStar, and APOGEE-2 Data}, {\it \apjs\/}, {\bf
  259}(2), 35.

\bibitem[{Abul Hayat} et~al.(2021){Abul Hayat}, {Harrington}, {Stein},
  {Luki{\'c}}, \& {Mustafa}]{Hayat-2021-Neurips}
{Abul Hayat}, M., {Harrington}, P., {Stein}, G., {Luki{\'c}}, Z., \& {Mustafa},
  M., 2021.
\newblock {Estimating Galactic Distances From Images Using Self-supervised
  Representation Learning}, {\it arXiv e-prints\/}, p. arXiv:2101.04293.

\bibitem[Arjovsky et~al.(2017)Arjovsky, Chintala, \&
  Bottou]{arjovsky2017wasserstein}
Arjovsky, M., Chintala, S., \& Bottou, L., 2017.
\newblock Wasserstein gan, {\it arXiv preprint arXiv:1701.07875\/}.

\bibitem[Belkin \& Niyogi(2003)]{Belkin:2003}
Belkin, M. \& Niyogi, P., 2003.
\newblock Laplacian eigenmaps for dimensionality reduction and data
  representation, {\it Neural Computation\/}, {\bf 15}, 1373--1396.

\bibitem[Bengio et~al.(2013)Bengio, Courville, \&
  Vincent]{bengio2013representation}
Bengio, Y., Courville, A., \& Vincent, P., 2013.
\newblock Representation learning: A review and new perspectives, {\it IEEE
  transactions on pattern analysis and machine intelligence\/}, {\bf 35}(8),
  1798--1828.

\bibitem[Biehl(2022)]{e08667bd2e844956b9d345b9e957634d}
Biehl, M., 2022.
\newblock {\it The Shallow and the Deep: A biased introduction to neural
  networks and old school machine learning\/}, Rijksuniversiteit Groningen.

\bibitem[Bishop(2006)]{bishop2006pattern}
Bishop, C.~M., 2006.
\newblock {\it Pattern recognition and machine learning\/}, Springer.

\bibitem[Brock et~al.(2018)Brock, Donahue, \& Simonyan]{brock2018large}
Brock, A., Donahue, J., \& Simonyan, K., 2018.
\newblock Large scale gan training for high fidelity natural image synthesis,
  in {\em International Conference on Learning Representations\/}.

\bibitem[{Bundy} et~al.(2015){Bundy}, {Bershady}, {Law}, {Yan}, {Drory},
  {MacDonald}, {Wake}, {Cherinka}, {S{\'a}nchez-Gallego}, {Weijmans}, {Thomas},
  {Tremonti}, {Masters}, {Coccato}, {Diamond-Stanic}, {Arag{\'o}n-Salamanca},
  {Avila-Reese}, {Badenes}, {Falc{\'o}n-Barroso}, {Belfiore}, {Bizyaev},
  {Blanc}, {Bland-Hawthorn}, {Blanton}, {Brownstein}, {Byler}, {Cappellari},
  {Conroy}, {Dutton}, {Emsellem}, {Etherington}, {Frinchaboy}, {Fu}, {Gunn},
  {Harding}, {Johnston}, {Kauffmann}, {Kinemuchi}, {Klaene}, {Knapen},
  {Leauthaud}, {Li}, {Lin}, {Maiolino}, {Malanushenko}, {Malanushenko}, {Mao},
  {Maraston}, {McDermid}, {Merrifield}, {Nichol}, {Oravetz}, {Pan}, {Parejko},
  {Sanchez}, {Schlegel}, {Simmons}, {Steele}, {Steinmetz}, {Thanjavur},
  {Thompson}, {Tinker}, {van den Bosch}, {Westfall}, {Wilkinson}, {Wright},
  {Xiao}, \& {Zhang}]{2015ApJ...798....7B}
{Bundy}, K., {Bershady}, M.~A., {Law}, D.~R., {Yan}, R., {Drory}, N.,
  {MacDonald}, N., {Wake}, D.~A., {Cherinka}, B., {S{\'a}nchez-Gallego}, J.~R.,
  {Weijmans}, A.-M., {Thomas}, D., {Tremonti}, C., {Masters}, K., {Coccato},
  L., {Diamond-Stanic}, A.~M., {Arag{\'o}n-Salamanca}, A., {Avila-Reese}, V.,
  {Badenes}, C., {Falc{\'o}n-Barroso}, J., {Belfiore}, F., {Bizyaev}, D.,
  {Blanc}, G.~A., {Bland-Hawthorn}, J., {Blanton}, M.~R., {Brownstein}, J.~R.,
  {Byler}, N., {Cappellari}, M., {Conroy}, C., {Dutton}, A.~A., {Emsellem}, E.,
  {Etherington}, J., {Frinchaboy}, P.~M., {Fu}, H., {Gunn}, J.~E., {Harding},
  P., {Johnston}, E.~J., {Kauffmann}, G., {Kinemuchi}, K., {Klaene}, M.~A.,
  {Knapen}, J.~H., {Leauthaud}, A., {Li}, C., {Lin}, L., {Maiolino}, R.,
  {Malanushenko}, V., {Malanushenko}, E., {Mao}, S., {Maraston}, C.,
  {McDermid}, R.~M., {Merrifield}, M.~R., {Nichol}, R.~C., {Oravetz}, D.,
  {Pan}, K., {Parejko}, J.~K., {Sanchez}, S.~F., {Schlegel}, D., {Simmons}, A.,
  {Steele}, O., {Steinmetz}, M., {Thanjavur}, K., {Thompson}, B.~A., {Tinker},
  J.~L., {van den Bosch}, R. C.~E., {Westfall}, K.~B., {Wilkinson}, D.,
  {Wright}, S., {Xiao}, T., \& {Zhang}, K., 2015.
\newblock {Overview of the SDSS-IV MaNGA Survey: Mapping nearby Galaxies at
  Apache Point Observatory}, {\it \apj\/}, {\bf 798}(1), 7.

\bibitem[Chechik et~al.(2010)Chechik, Sharma, Shalit, \& Bengio]{OASIS-2010}
Chechik, G., Sharma, V., Shalit, U., \& Bengio, S., 2010.
\newblock Large scale online learning of image similarity through ranking, {\it
  Journal of Machine Learning Research\/}, {\bf 11}(36), 1109--1135.

\bibitem[Chen et~al.(2018{\natexlab{a}})Chen, Li, Grosse, \&
  Duvenaud]{chen2018isolating}
Chen, R.~T., Li, X., Grosse, R.~B., \& Duvenaud, D.~K., 2018{\natexlab{a}}.
\newblock Isolating sources of disentanglement in variational autoencoders, in
  {\em Advances in Neural Information Processing Systems\/}, pp. 7512--7523.

\bibitem[Chen et~al.(2018{\natexlab{b}})Chen, Rubanova, Bettencourt, \&
  Duvenaud]{chen2018neural}
Chen, R.~T., Rubanova, Y., Bettencourt, J., \& Duvenaud, D.,
  2018{\natexlab{b}}.
\newblock Neural ordinary differential equations, in {\em Advances in Neural
  Information Processing Systems\/}, pp. 6571--6583.

\bibitem[Chen et~al.(2021)Chen, Fan, Agrawal, \& Duvenaud]{chen2021conditional}
Chen, R.~T., Fan, J., Agrawal, S., \& Duvenaud, D., 2021.
\newblock Conditional diffusion models for continuous and discrete outcomes, in
  {\em International Conference on Machine Learning\/}, pp. 1716--1726.

\bibitem[Chen et~al.(2020{\natexlab{a}})Chen, Kornblith, Norouzi, \&
  Hinton]{SimCLR1}
Chen, T., Kornblith, S., Norouzi, M., \& Hinton, G., 2020{\natexlab{a}}.
\newblock A simple framework for contrastive learning of visual
  representations.

\bibitem[Chen \& He(2020)]{SimSiam}
Chen, X. \& He, K., 2020.
\newblock Exploring simple siamese representation learning.

\bibitem[Chen et~al.(2020{\natexlab{b}})Chen, Fan, Girshick, \& He]{MoCo2}
Chen, X., Fan, H., Girshick, R., \& He, K., 2020{\natexlab{b}}.
\newblock Improved baselines with momentum contrastive learning.

\bibitem[Chen \& Bach(2021)]{chen2021stochastic}
Chen, X.-J. \& Bach, F., 2021.
\newblock Stochastic differential equations as diffusions: Reinterpreting flow
  models as continuous-time markov chains, {\it International Conference on
  Learning Representations\/}.

\bibitem[Chopra et~al.(2005)Chopra, Hadsell, \& LeCun]{Chopra-2005}
Chopra, S., Hadsell, R., \& LeCun, Y., 2005.
\newblock Learning a similarity metric discriminatively, with application to
  face verification, in {\em 2005 IEEE Computer Society Conference on Computer
  Vision and Pattern Recognition (CVPR'05)\/}, vol.~1, pp. 539--546 vol. 1.

\bibitem[{{\'C}iprijanovi{\'c}} et~al.(2020){{\'C}iprijanovi{\'c}}, {Snyder},
  {Nord}, \& {Peek}]{2020A&C....3200390C}
{{\'C}iprijanovi{\'c}}, A., {Snyder}, G.~F., {Nord}, B., \& {Peek}, J.~E.~G.,
  2020.
\newblock {DeepMerge: Classifying high-redshift merging galaxies with deep
  neural networks}, {\it Astronomy and Computing\/}, {\bf 32}, 100390.

\bibitem[{{\'C}iprijanovi{\'c}} et~al.(2023){{\'C}iprijanovi{\'c}}, {Lewis},
  {Pedro}, {Madireddy}, {Nord}, {Perdue}, \& {Wild}]{2023arXiv230202005C}
{{\'C}iprijanovi{\'c}}, A., {Lewis}, A., {Pedro}, K., {Madireddy}, S., {Nord},
  B., {Perdue}, G.~N., \& {Wild}, S.~M., 2023.
\newblock {DeepAstroUDA: Semi-Supervised Universal Domain Adaptation for
  Cross-Survey Galaxy Morphology Classification and Anomaly Detection}, {\it
  arXiv e-prints\/}, p. arXiv:2302.02005.

\bibitem[{Dey} et~al.(2019){Dey}, {Schlegel}, {Lang}, {Blum}, {Burleigh},
  {Fan}, {Findlay}, {Finkbeiner}, {Herrera}, {Juneau}, {Landriau}, {Levi},
  {McGreer}, {Meisner}, {Myers}, {Moustakas}, {Nugent}, {Patej}, {Schlafly},
  {Walker}, {Valdes}, {Weaver}, {Y{\`e}che}, {Zou}, {Zhou}, {Abareshi},
  {Abbott}, {Abolfathi}, {Aguilera}, {Alam}, {Allen}, {Alvarez}, {Annis},
  {Ansarinejad}, {Aubert}, {Beechert}, {Bell}, {BenZvi}, {Beutler}, {Bielby},
  {Bolton}, {Brice{\~n}o}, {Buckley-Geer}, {Butler}, {Calamida}, {Carlberg},
  {Carter}, {Casas}, {Castander}, {Choi}, {Comparat}, {Cukanovaite}, {Delubac},
  {DeVries}, {Dey}, {Dhungana}, {Dickinson}, {Ding}, {Donaldson}, {Duan},
  {Duckworth}, {Eftekharzadeh}, {Eisenstein}, {Etourneau}, {Fagrelius},
  {Farihi}, {Fitzpatrick}, {Font-Ribera}, {Fulmer}, {G{\"a}nsicke},
  {Gaztanaga}, {George}, {Gerdes}, {Gontcho}, {Gorgoni}, {Green}, {Guy},
  {Harmer}, {Hernandez}, {Honscheid}, {Huang}, {James}, {Jannuzi}, {Jiang},
  {Joyce}, {Karcher}, {Karkar}, {Kehoe}, {Kneib}, {Kueter-Young}, {Lan},
  {Lauer}, {Le Guillou}, {Le Van Suu}, {Lee}, {Lesser}, {Perreault Levasseur},
  {Li}, {Mann}, {Marshall}, {Mart{\'\i}nez-V{\'a}zquez}, {Martini}, {du Mas des
  Bourboux}, {McManus}, {Meier}, {M{\'e}nard}, {Metcalfe},
  {Mu{\~n}oz-Guti{\'e}rrez}, {Najita}, {Napier}, {Narayan}, {Newman}, {Nie},
  {Nord}, {Norman}, {Olsen}, {Paat}, {Palanque-Delabrouille}, {Peng},
  {Poppett}, {Poremba}, {Prakash}, {Rabinowitz}, {Raichoor}, {Rezaie},
  {Robertson}, {Roe}, {Ross}, {Ross}, {Rudnick}, {Safonova}, {Saha},
  {S{\'a}nchez}, {Savary}, {Schweiker}, {Scott}, {Seo}, {Shan}, {Silva},
  {Slepian}, {Soto}, {Sprayberry}, {Staten}, {Stillman}, {Stupak}, {Summers},
  {Sien Tie}, {Tirado}, {Vargas-Maga{\~n}a}, {Vivas}, {Wechsler}, {Williams},
  {Yang}, {Yang}, {Yapici}, {Zaritsky}, {Zenteno}, {Zhang}, {Zhang}, {Zhou}, \&
  {Zhou}]{2019AJ....157..168D}
{Dey}, A., {Schlegel}, D.~J., {Lang}, D., {Blum}, R., {Burleigh}, K., {Fan},
  X., {Findlay}, J.~R., {Finkbeiner}, D., {Herrera}, D., {Juneau}, S.,
  {Landriau}, M., {Levi}, M., {McGreer}, I., {Meisner}, A., {Myers}, A.~D.,
  {Moustakas}, J., {Nugent}, P., {Patej}, A., {Schlafly}, E.~F., {Walker},
  A.~R., {Valdes}, F., {Weaver}, B.~A., {Y{\`e}che}, C., {Zou}, H., {Zhou}, X.,
  {Abareshi}, B., {Abbott}, T.~M.~C., {Abolfathi}, B., {Aguilera}, C., {Alam},
  S., {Allen}, L., {Alvarez}, A., {Annis}, J., {Ansarinejad}, B., {Aubert}, M.,
  {Beechert}, J., {Bell}, E.~F., {BenZvi}, S.~Y., {Beutler}, F., {Bielby},
  R.~M., {Bolton}, A.~S., {Brice{\~n}o}, C., {Buckley-Geer}, E.~J., {Butler},
  K., {Calamida}, A., {Carlberg}, R.~G., {Carter}, P., {Casas}, R.,
  {Castander}, F.~J., {Choi}, Y., {Comparat}, J., {Cukanovaite}, E., {Delubac},
  T., {DeVries}, K., {Dey}, S., {Dhungana}, G., {Dickinson}, M., {Ding}, Z.,
  {Donaldson}, J.~B., {Duan}, Y., {Duckworth}, C.~J., {Eftekharzadeh}, S.,
  {Eisenstein}, D.~J., {Etourneau}, T., {Fagrelius}, P.~A., {Farihi}, J.,
  {Fitzpatrick}, M., {Font-Ribera}, A., {Fulmer}, L., {G{\"a}nsicke}, B.~T.,
  {Gaztanaga}, E., {George}, K., {Gerdes}, D.~W., {Gontcho}, S. G.~A.,
  {Gorgoni}, C., {Green}, G., {Guy}, J., {Harmer}, D., {Hernandez}, M.,
  {Honscheid}, K., {Huang}, L.~W., {James}, D.~J., {Jannuzi}, B.~T., {Jiang},
  L., {Joyce}, R., {Karcher}, A., {Karkar}, S., {Kehoe}, R., {Kneib}, J.-P.,
  {Kueter-Young}, A., {Lan}, T.-W., {Lauer}, T.~R., {Le Guillou}, L., {Le Van
  Suu}, A., {Lee}, J.~H., {Lesser}, M., {Perreault Levasseur}, L., {Li}, T.~S.,
  {Mann}, J.~L., {Marshall}, R., {Mart{\'\i}nez-V{\'a}zquez}, C.~E., {Martini},
  P., {du Mas des Bourboux}, H., {McManus}, S., {Meier}, T.~G., {M{\'e}nard},
  B., {Metcalfe}, N., {Mu{\~n}oz-Guti{\'e}rrez}, A., {Najita}, J., {Napier},
  K., {Narayan}, G., {Newman}, J.~A., {Nie}, J., {Nord}, B., {Norman}, D.~J.,
  {Olsen}, K. A.~G., {Paat}, A., {Palanque-Delabrouille}, N., {Peng}, X.,
  {Poppett}, C.~L., {Poremba}, M.~R., {Prakash}, A., {Rabinowitz}, D.,
  {Raichoor}, A., {Rezaie}, M., {Robertson}, A.~N., {Roe}, N.~A., {Ross},
  A.~J., {Ross}, N.~P., {Rudnick}, G., {Safonova}, S., {Saha}, A.,
  {S{\'a}nchez}, F.~J., {Savary}, E., {Schweiker}, H., {Scott}, A., {Seo},
  H.-J., {Shan}, H., {Silva}, D.~R., {Slepian}, Z., {Soto}, C., {Sprayberry},
  D., {Staten}, R., {Stillman}, C.~M., {Stupak}, R.~J., {Summers}, D.~L., {Sien
  Tie}, S., {Tirado}, H., {Vargas-Maga{\~n}a}, M., {Vivas}, A.~K., {Wechsler},
  R.~H., {Williams}, D., {Yang}, J., {Yang}, Q., {Yapici}, T., {Zaritsky}, D.,
  {Zenteno}, A., {Zhang}, K., {Zhang}, T., {Zhou}, R., \& {Zhou}, Z., 2019.
\newblock {Overview of the DESI Legacy Imaging Surveys}, {\it \aj\/}, {\bf
  157}(5), 168.

\bibitem[Dinh et~al.(2017)Dinh, Sohl-Dickstein, \& Bengio]{dinh2016density}
Dinh, L., Sohl-Dickstein, J., \& Bengio, S., 2017.
\newblock Density estimation using real nvp, in {\em International Conference
  on Learning Representations\/}.

\bibitem[Doersch(2016)]{doersch2016tutorial}
Doersch, C., 2016.
\newblock Tutorial on variational autoencoders, {\it arXiv preprint
  arXiv:1606.05908\/}.

\bibitem[{Dom{\'\i}nguez S{\'a}nchez} et~al.(2019){Dom{\'\i}nguez S{\'a}nchez},
  {Huertas-Company}, {Bernardi}, {Kaviraj}, {Fischer}, {Abbott}, {Abdalla},
  {Annis}, {Avila}, {Brooks}, {Buckley-Geer}, {Carnero Rosell}, {Carrasco
  Kind}, {Carretero}, {Cunha}, {D'Andrea}, {da Costa}, {Davis}, {De Vicente},
  {Doel}, {Evrard}, {Fosalba}, {Frieman}, {Garc{\'\i}a-Bellido}, {Gaztanaga},
  {Gerdes}, {Gruen}, {Gruendl}, {Gschwend}, {Gutierrez}, {Hartley},
  {Hollowood}, {Honscheid}, {Hoyle}, {James}, {Kuehn}, {Kuropatkin}, {Lahav},
  {Maia}, {March}, {Melchior}, {Menanteau}, {Miquel}, {Nord}, {Plazas},
  {Sanchez}, {Scarpine}, {Schindler}, {Schubnell}, {Smith}, {Smith},
  {Soares-Santos}, {Sobreira}, {Suchyta}, {Swanson}, {Tarle}, {Thomas},
  {Walker}, \& {Zuntz}]{2019MNRAS.484...93D}
{Dom{\'\i}nguez S{\'a}nchez}, H., {Huertas-Company}, M., {Bernardi}, M.,
  {Kaviraj}, S., {Fischer}, J.~L., {Abbott}, T.~M.~C., {Abdalla}, F.~B.,
  {Annis}, J., {Avila}, S., {Brooks}, D., {Buckley-Geer}, E., {Carnero Rosell},
  A., {Carrasco Kind}, M., {Carretero}, J., {Cunha}, C.~E., {D'Andrea}, C.~B.,
  {da Costa}, L.~N., {Davis}, C., {De Vicente}, J., {Doel}, P., {Evrard},
  A.~E., {Fosalba}, P., {Frieman}, J., {Garc{\'\i}a-Bellido}, J., {Gaztanaga},
  E., {Gerdes}, D.~W., {Gruen}, D., {Gruendl}, R.~A., {Gschwend}, J.,
  {Gutierrez}, G., {Hartley}, W.~G., {Hollowood}, D.~L., {Honscheid}, K.,
  {Hoyle}, B., {James}, D.~J., {Kuehn}, K., {Kuropatkin}, N., {Lahav}, O.,
  {Maia}, M.~A.~G., {March}, M., {Melchior}, P., {Menanteau}, F., {Miquel}, R.,
  {Nord}, B., {Plazas}, A.~A., {Sanchez}, E., {Scarpine}, V., {Schindler}, R.,
  {Schubnell}, M., {Smith}, M., {Smith}, R.~C., {Soares-Santos}, M.,
  {Sobreira}, F., {Suchyta}, E., {Swanson}, M.~E.~C., {Tarle}, G., {Thomas},
  D., {Walker}, A.~R., \& {Zuntz}, J., 2019.
\newblock {Transfer learning for galaxy morphology from one survey to another},
  {\it \mnras\/}, {\bf 484}(1), 93--100.

\bibitem[Donoho \& Grimes(2003)]{Hess-maps}
Donoho, D.~L. \& Grimes, C., 2003.
\newblock Hessian eigenmaps: Locally linear embedding techniques for
  high-dimensional data, {\it Proceedings of the National Academy of
  Sciences\/}, {\bf 100}(10), 5591--5596.

\bibitem[{Doorenbos} et~al.(2022){Doorenbos}, {Cavuoti}, {Longo}, {Brescia},
  {Sznitman}, \& {M{\'a}rquez-Neila}]{Doorenbos-2022}
{Doorenbos}, L., {Cavuoti}, S., {Longo}, G., {Brescia}, M., {Sznitman}, R., \&
  {M{\'a}rquez-Neila}, P., 2022.
\newblock {Generating astronomical spectra from photometry with conditional
  diffusion models}, {\it arXiv e-prints\/}, p. arXiv:2211.05556.

\bibitem[{Finkelstein} et~al.(2022){Finkelstein}, {Bagley}, {Haro},
  {Dickinson}, {Ferguson}, {Kartaltepe}, {Papovich}, {Burgarella}, {Kocevski},
  {Huertas-Company}, {Iyer}, {Koekemoer}, {Larson}, {P{\'e}rez-Gonz{\'a}lez},
  {Rose}, {Tacchella}, {Wilkins}, {Chworowsky}, {Medrano}, {Morales},
  {Somerville}, {Yung}, {Fontana}, {Giavalisco}, {Grazian}, {Grogin}, {Kewley},
  {Kirkpatrick}, {Kurczynski}, {Lotz}, {Pentericci}, {Pirzkal}, {Ravindranath},
  {Ryan}, {Trump}, {Yang}, {Almaini}, {Amor{\'\i}n}, {Annunziatella},
  {Backhaus}, {Barro}, {Behroozi}, {Bell}, {Bhatawdekar}, {Bisigello}, {Bromm},
  {Buat}, {Buitrago}, {Calabr{\`o}}, {Casey}, {Castellano}, {Ch{\'a}vez Ortiz},
  {Ciesla}, {Cleri}, {Cohen}, {Cole}, {Cooke}, {Cooper}, {Cooray}, {Costantin},
  {Cox}, {Croton}, {Daddi}, {Dav{\'e}}, {de La Vega}, {Dekel}, {Elbaz},
  {Estrada-Carpenter}, {Faber}, {Fern{\'a}ndez}, {Finkelstein}, {Freundlich},
  {Fujimoto}, {Garc{\'\i}a-Argum{\'a}nez}, {Gardner}, {Gawiser},
  {G{\'o}mez-Guijarro}, {Guo}, {Hamblin}, {Hamilton}, {Hathi}, {Holwerda},
  {Hirschmann}, {Hutchison}, {Jaskot}, {Jha}, {Jogee}, {Juneau}, {Jung},
  {Kassin}, {Le Bail}, {Leung}, {Lucas}, {Magnelli}, {Mantha}, {Matharu},
  {McGrath}, {McIntosh}, {Merlin}, {Mobasher}, {Newman}, {Nicholls}, {Pandya},
  {Rafelski}, {Ronayne}, {Santini}, {Seill{\'e}}, {Shah}, {Shen}, {Simons},
  {Snyder}, {Stanway}, {Straughn}, {Teplitz}, {Vanderhoof}, {Vega-Ferrero},
  {Wang}, {Weiner}, {Willmer}, {Wuyts}, {Zavala}, \& {CEERS
  Team}]{2022ApJ...940L..55F}
{Finkelstein}, S.~L., {Bagley}, M.~B., {Haro}, P.~A., {Dickinson}, M.,
  {Ferguson}, H.~C., {Kartaltepe}, J.~S., {Papovich}, C., {Burgarella}, D.,
  {Kocevski}, D.~D., {Huertas-Company}, M., {Iyer}, K.~G., {Koekemoer}, A.~M.,
  {Larson}, R.~L., {P{\'e}rez-Gonz{\'a}lez}, P.~G., {Rose}, C., {Tacchella},
  S., {Wilkins}, S.~M., {Chworowsky}, K., {Medrano}, A., {Morales}, A.~M.,
  {Somerville}, R.~S., {Yung}, L.~Y.~A., {Fontana}, A., {Giavalisco}, M.,
  {Grazian}, A., {Grogin}, N.~A., {Kewley}, L.~J., {Kirkpatrick}, A.,
  {Kurczynski}, P., {Lotz}, J.~M., {Pentericci}, L., {Pirzkal}, N.,
  {Ravindranath}, S., {Ryan}, R.~E., {Trump}, J.~R., {Yang}, G., {Almaini}, O.,
  {Amor{\'\i}n}, R.~O., {Annunziatella}, M., {Backhaus}, B.~E., {Barro}, G.,
  {Behroozi}, P., {Bell}, E.~F., {Bhatawdekar}, R., {Bisigello}, L., {Bromm},
  V., {Buat}, V., {Buitrago}, F., {Calabr{\`o}}, A., {Casey}, C.~M.,
  {Castellano}, M., {Ch{\'a}vez Ortiz}, {\'O}.~A., {Ciesla}, L., {Cleri},
  N.~J., {Cohen}, S.~H., {Cole}, J.~W., {Cooke}, K.~C., {Cooper}, M.~C.,
  {Cooray}, A.~R., {Costantin}, L., {Cox}, I.~G., {Croton}, D., {Daddi}, E.,
  {Dav{\'e}}, R., {de La Vega}, A., {Dekel}, A., {Elbaz}, D.,
  {Estrada-Carpenter}, V., {Faber}, S.~M., {Fern{\'a}ndez}, V., {Finkelstein},
  K.~D., {Freundlich}, J., {Fujimoto}, S., {Garc{\'\i}a-Argum{\'a}nez}, {\'A}.,
  {Gardner}, J.~P., {Gawiser}, E., {G{\'o}mez-Guijarro}, C., {Guo}, Y.,
  {Hamblin}, K., {Hamilton}, T.~S., {Hathi}, N.~P., {Holwerda}, B.~W.,
  {Hirschmann}, M., {Hutchison}, T.~A., {Jaskot}, A.~E., {Jha}, S.~W., {Jogee},
  S., {Juneau}, S., {Jung}, I., {Kassin}, S.~A., {Le Bail}, A., {Leung}, G.
  C.~K., {Lucas}, R.~A., {Magnelli}, B., {Mantha}, K.~B., {Matharu}, J.,
  {McGrath}, E.~J., {McIntosh}, D.~H., {Merlin}, E., {Mobasher}, B., {Newman},
  J.~A., {Nicholls}, D.~C., {Pandya}, V., {Rafelski}, M., {Ronayne}, K.,
  {Santini}, P., {Seill{\'e}}, L.-M., {Shah}, E.~A., {Shen}, L., {Simons},
  R.~C., {Snyder}, G.~F., {Stanway}, E.~R., {Straughn}, A.~N., {Teplitz},
  H.~I., {Vanderhoof}, B.~N., {Vega-Ferrero}, J., {Wang}, W., {Weiner}, B.~J.,
  {Willmer}, C. N.~A., {Wuyts}, S., {Zavala}, J.~A., \& {CEERS Team}, 2022.
\newblock {A Long Time Ago in a Galaxy Far, Far Away: A Candidate z 12 Galaxy
  in Early JWST CEERS Imaging}, {\it \apjl\/}, {\bf 940}(2), L55.

\bibitem[Goodfellow et~al.(2014)Goodfellow, Pouget-Abadie, Mirza, Xu,
  Warde-Farley, Ozair, Courville, \& Bengio]{goodfellow2014generative}
Goodfellow, I., Pouget-Abadie, J., Mirza, M., Xu, B., Warde-Farley, D., Ozair,
  S., Courville, A., \& Bengio, Y., 2014.
\newblock Generative adversarial nets, in {\em Advances in Neural Information
  Processing Systems\/}, pp. 2672--2680.

\bibitem[Grathwohl et~al.(2019)Grathwohl, Chen, Betterncourt, Sutskever, \&
  Duvenaud]{grathwohl2018ffjord}
Grathwohl, W., Chen, R.~T., Betterncourt, J., Sutskever, I., \& Duvenaud, D.,
  2019.
\newblock Ffjord: Free-form continuous dynamics for scalable reversible
  generative models, in {\em International Conference on Learning
  Representations\/}.

\bibitem[Grathwohl et~al.(2021)Grathwohl, Chen, Jacobsen, \&
  Duvenaud]{grathwohl2021your}
Grathwohl, W., Chen, R.~T., Jacobsen, J.-H., \& Duvenaud, D., 2021.
\newblock Your classifier is secretly an energy based model and you should
  treat it like one, {\it International Conference on Learning
  Representations\/}.

\bibitem[Grill et~al.(2020)Grill, Strub, Altché, Tallec, Richemond,
  Buchatskaya, Doersch, Pires, Guo, Azar, Piot, Kavukcuoglu, Munos, \&
  Valko]{BYOL}
Grill, J.-B., Strub, F., Altché, F., Tallec, C., Richemond, P.~H.,
  Buchatskaya, E., Doersch, C., Pires, B.~A., Guo, Z.~D., Azar, M.~G., Piot,
  B., Kavukcuoglu, K., Munos, R., \& Valko, M., 2020.
\newblock Bootstrap your own latent: A new approach to self-supervised
  learning.

\bibitem[{Guo} et~al.(2022){Guo}, {Liu}, {Qiu}, {Luo}, {Jiang}, {Shi}, {Li}, \&
  {Wang}]{Guo-2022}
{Guo}, X., {Liu}, C., {Qiu}, B., {Luo}, A.~l., {Jiang}, X., {Shi}, J., {Li},
  X., \& {Wang}, L., 2022.
\newblock {Unsupervised clustering and analysis of WISE spiral galaxies}, {\it
  \mnras\/}, {\bf 517}(2), 1837--1848.

\bibitem[Hadsell et~al.(2006)Hadsell, Chopra, \& LeCun]{Hadsell-2006}
Hadsell, R., Chopra, S., \& LeCun, Y., 2006.
\newblock Dimensionality reduction by learning an invariant mapping, in {\em
  2006 IEEE Computer Society Conference on Computer Vision and Pattern
  Recognition (CVPR'06)\/}, vol.~2, pp. 1735--1742.

\bibitem[{Hayat} et~al.(2021){Hayat}, {Stein}, {Harrington}, {Luki{\'c}}, \&
  {Mustafa}]{Hayat-2021}
{Hayat}, M.~A., {Stein}, G., {Harrington}, P., {Luki{\'c}}, Z., \& {Mustafa},
  M., 2021.
\newblock {Self-supervised Representation Learning for Astronomical Images},
  {\it \apjl\/}, {\bf 911}(2), L33.

\bibitem[He et~al.(2019)He, Fan, Wu, Xie, \& Girshick]{MoCo1}
He, K., Fan, H., Wu, Y., Xie, S., \& Girshick, R., 2019.
\newblock Momentum contrast for unsupervised visual representation learning.

\bibitem[Higgins et~al.(2017)Higgins, Matthey, Pal, Burgess, Glorot, Botvinick,
  Mohamed, \& Lerchner]{higgins2017beta}
Higgins, I., Matthey, L., Pal, A., Burgess, C., Glorot, X., Botvinick, M.,
  Mohamed, S., \& Lerchner, A., 2017.
\newblock Beta-vae: Learning basic visual concepts with a constrained
  variational framework, in {\em International Conference on Learning
  Representations\/}.

\bibitem[Hinton \& Salakhutdinov(2006)]{hinton2006reducing}
Hinton, G.~E. \& Salakhutdinov, R.~R., 2006.
\newblock Reducing the dimensionality of data with neural networks, {\it
  science\/}, {\bf 313}(5786), 504--507.

\bibitem[Hjelm et~al.(2019)Hjelm, Fedorov, Lavoie-Marchildon, Grewal, Bachman,
  Trischler, \& Bengio]{hjelm2019learning}
Hjelm, R.~D., Fedorov, A., Lavoie-Marchildon, S., Grewal, K., Bachman, P.,
  Trischler, A., \& Bengio, Y., 2019.
\newblock Learning deep representations by mutual information estimation and
  maximization.

\bibitem[Ho et~al.(2020)Ho, Chen, Srinivas, Dinh, Abbeel, \&
  Song]{ho2020denoising}
Ho, J., Chen, X., Srinivas, A., Dinh, L., Abbeel, P., \& Song, J., 2020.
\newblock Denoising diffusion probabilistic models, {\it Advances in Neural
  Information Processing Systems\/}, {\bf 33}.

\bibitem[{Huertas-Company} \& {Lanusse}(2023)]{2023PASA...40....1H}
{Huertas-Company}, M. \& {Lanusse}, F., 2023.
\newblock {The Dawes Review 10: The impact of deep learning for the analysis of
  galaxy surveys}, {\it \pasa\/}, {\bf 40}, e001.

\bibitem[Karras et~al.(2019)Karras, Laine, \& Aila]{karras2019style}
Karras, T., Laine, S., \& Aila, T., 2019.
\newblock A style-based generator architecture for generative adversarial
  networks, in {\em Conference on Computer Vision and Pattern Recognition\/},
  pp. 4401--4410.

\bibitem[Khosla et~al.(2020)Khosla, Teterwak, Wang, Sarna, Tian, Isola,
  Maschinot, Liu, \& Krishnan]{SupCon-2020}
Khosla, P., Teterwak, P., Wang, C., Sarna, A., Tian, Y., Isola, P., Maschinot,
  A., Liu, C., \& Krishnan, D., 2020.
\newblock Supervised contrastive learning, in {\em Advances in Neural
  Information Processing Systems\/}, vol.~33, pp. 18661--18673, Curran
  Associates, Inc.

\bibitem[Kingma \& Welling(2013)]{kingma2013auto}
Kingma, D.~P. \& Welling, M., 2013.
\newblock Auto-encoding variational bayes, {\it arXiv preprint
  arXiv:1312.6114\/}.

\bibitem[Kingma et~al.(2016)Kingma, Salimans, Jozefowicz, Chen, Sutskever, \&
  Welling]{kingma2016improved}
Kingma, D.~P., Salimans, T., Jozefowicz, R., Chen, X., Sutskever, I., \&
  Welling, M., 2016.
\newblock Improved variational inference with inverse autoregressive flow, in
  {\em Advances in Neural Information Processing Systems\/}, pp. 4743--4751.

\bibitem[{Lamdouar} et~al.(2022){Lamdouar}, {Sundaresan}, {Jungbluth}, {Boro
  Saikia}, {Camarata}, {Miles}, {Scoczynski}, {Stone}, {Sarah},
  {Mu{\~n}oz-Jaramillo}, {Narock}, \& {Szabo}]{Lamdouar-2022}
{Lamdouar}, H., {Sundaresan}, S., {Jungbluth}, A., {Boro Saikia}, S.,
  {Camarata}, A.~J., {Miles}, N., {Scoczynski}, M., {Stone}, M., {Sarah}, A.,
  {Mu{\~n}oz-Jaramillo}, A., {Narock}, A., \& {Szabo}, A., 2022.
\newblock {Deep-SWIM: A few-shot learning approach to classify Solar WInd
  Magnetic field structures}, {\it arXiv e-prints\/}, p. arXiv:2203.01184.

\bibitem[Le-Khac et~al.(2020)Le-Khac, Healy, \& Smeaton]{Le_Khac-2020}
Le-Khac, P.~H., Healy, G., \& Smeaton, A.~F., 2020.
\newblock Contrastive representation learning: A framework and review, {\it
  {IEEE} Access\/}, {\bf 8}, 193907--193934.

\bibitem[Li et~al.(2021)Li, Shen, Zhang, Li, Van Den~Hengel, \&
  Wang]{li2021survey}
Li, H., Shen, C., Zhang, R., Li, G., Van Den~Hengel, A., \& Wang, Q., 2021.
\newblock A survey on deep domain adaptation, {\it IEEE Transactions on Neural
  Networks and Learning Systems\/}, {\bf 32}(3), 660--699.

\bibitem[{Li} et~al.(2021){Li}, {Li}, {Shi}, \& {Yu}]{2021arXiv210409415L}
{Li}, J., {Li}, G., {Shi}, Y., \& {Yu}, Y., 2021.
\newblock {Cross-Domain Adaptive Clustering for Semi-Supervised Domain
  Adaptation}, {\it arXiv e-prints\/}, p. arXiv:2104.09415.

\bibitem[Liu et~al.(2021)Liu, Wang, Yang, Cheng, Chen, \& Gao]{liu2021learning}
Liu, T., Wang, W., Yang, Y., Cheng, Y., Chen, H., \& Gao, J., 2021.
\newblock Learning langevin dynamics with diffusion models, in {\em
  International Conference on Machine Learning\/}, pp. 6444--6454.

\bibitem[McInnes et~al.(2018)McInnes, Healy, \& Melville]{umap}
McInnes, L., Healy, J., \& Melville, J., 2018.
\newblock Umap: Uniform manifold approximation and projection for dimension
  reduction.

\bibitem[{Mercea} et~al.(2023){Mercea}, {Paraschiv}, {Lacatus}, {Marginean}, \&
  {Besliu-Ionescu}]{sunquake_detect}
{Mercea}, V., {Paraschiv}, A.~R., {Lacatus}, D.~A., {Marginean}, A., \&
  {Besliu-Ionescu}, D., 2023.
\newblock {A Machine Learning Enhanced Approach for Automated Sunquake
  Detection in Acoustic Emission Maps}, {\it \solphys\/}, {\bf 298}(1), 4.

\bibitem[Misra \& van~der Maaten(2019)]{Misra-2019}
Misra, I. \& van~der Maaten, L., 2019.
\newblock Self-supervised learning of pretext-invariant representations.

\bibitem[Murphy(2022)]{pml1Book}
Murphy, K.~P., 2022.
\newblock {\it Probabilistic Machine Learning: An introduction\/}, MIT Press.

\bibitem[Oord et~al.(2018)Oord, Li, \& Vinyals]{CPC}
Oord, A. v.~d., Li, Y., \& Vinyals, O., 2018.
\newblock Representation learning with contrastive predictive coding.

\bibitem[Papamakarios et~al.(2017)Papamakarios, Pavlakou, \&
  Murray]{papamakarios2017masked}
Papamakarios, G., Pavlakou, T., \& Murray, I., 2017.
\newblock Masked autoregressive flow for density estimation, in {\em Advances
  in Neural Information Processing Systems\/}, pp. 2338--2347.

\bibitem[Pearson(1901)]{PCA-1901}
Pearson, K., 1901.
\newblock Liii. on lines and planes of closest fit to systems of points in
  space, {\it The London, Edinburgh, and Dublin Philosophical Magazine and
  Journal of Science\/}, {\bf 2}(11), 559--572.

\bibitem[Radford et~al.(2016)Radford, Metz, \&
  Chintala]{radford2015unsupervised}
Radford, A., Metz, L., \& Chintala, S., 2016.
\newblock Unsupervised representation learning with deep convolutional
  generative adversarial networks, in {\em International Conference on Learning
  Representations\/}.

\bibitem[{Radford} et~al.(2021){Radford}, {Kim}, {Hallacy}, {Ramesh}, {Goh},
  {Agarwal}, {Sastry}, {Askell}, {Mishkin}, {Clark}, {Krueger}, \&
  {Sutskever}]{2021arXiv210300020R}
{Radford}, A., {Kim}, J.~W., {Hallacy}, C., {Ramesh}, A., {Goh}, G., {Agarwal},
  S., {Sastry}, G., {Askell}, A., {Mishkin}, P., {Clark}, J., {Krueger}, G., \&
  {Sutskever}, I., 2021.
\newblock {Learning Transferable Visual Models From Natural Language
  Supervision}, {\it arXiv e-prints\/}, p. arXiv:2103.00020.

\bibitem[{Ramesh} et~al.(2021){Ramesh}, {Pavlov}, {Goh}, {Gray}, {Voss},
  {Radford}, {Chen}, \& {Sutskever}]{2021arXiv210212092R}
{Ramesh}, A., {Pavlov}, M., {Goh}, G., {Gray}, S., {Voss}, C., {Radford}, A.,
  {Chen}, M., \& {Sutskever}, I., 2021.
\newblock {Zero-Shot Text-to-Image Generation}, {\it arXiv e-prints\/}, p.
  arXiv:2102.12092.

\bibitem[Rezende \& Mohamed(2015)]{rezende2015variational}
Rezende, D.~J. \& Mohamed, S., 2015.
\newblock Variational inference with normalizing flows, in {\em International
  Conference on Machine Learning\/}, pp. 1530--1538.

\bibitem[Rezende et~al.(2014)Rezende, Mohamed, \&
  Wierstra]{rezende2014stochastic}
Rezende, D.~J., Mohamed, S., \& Wierstra, D., 2014.
\newblock Stochastic backpropagation and approximate inference in deep
  generative models, {\it arXiv preprint arXiv:1401.4082\/}.

\bibitem[Roweis \& Saul(2000)]{LLE}
Roweis, S.~T. \& Saul, L.~K., 2000.
\newblock Nonlinear dimensionality reduction by locally linear embedding, {\it
  Science\/}, {\bf 290}(5500), 2323--2326.

\bibitem[{Saito} et~al.(2020){Saito}, {Kim}, {Sclaroff}, \&
  {Saenko}]{2020arXiv200207953S}
{Saito}, K., {Kim}, D., {Sclaroff}, S., \& {Saenko}, K., 2020.
\newblock {Universal Domain Adaptation through Self Supervision}, {\it arXiv
  e-prints\/}, p. arXiv:2002.07953.

\bibitem[Salimans et~al.(2016)Salimans, Goodfellow, Zaremba, Cheung, Radford,
  \& Chen]{salimans2016improved}
Salimans, T., Goodfellow, I., Zaremba, W., Cheung, V., Radford, A., \& Chen,
  X., 2016.
\newblock Improved techniques for training gans, in {\em Advances in Neural
  Information Processing Systems\/}, pp. 2234--2242.

\bibitem[{Sarmiento} et~al.(2021){Sarmiento}, {Huertas-Company}, {Knapen},
  {S{\'a}nchez}, {Dom{\'\i}nguez S{\'a}nchez}, {Drory}, \&
  {Falc{\'o}n-Barroso}]{Sarmiento-2021}
{Sarmiento}, R., {Huertas-Company}, M., {Knapen}, J.~H., {S{\'a}nchez}, S.~F.,
  {Dom{\'\i}nguez S{\'a}nchez}, H., {Drory}, N., \& {Falc{\'o}n-Barroso}, J.,
  2021.
\newblock {Capturing the Physics of MaNGA Galaxies with Self-supervised Machine
  Learning}, {\it \apj\/}, {\bf 921}(2), 177.

\bibitem[Sch{\"o}lkopf et~al.(1997)Sch{\"o}lkopf, Smola, \&
  M{\"u}ller]{KernelPCA}
Sch{\"o}lkopf, B., Smola, A., \& M{\"u}ller, K.-R., 1997.
\newblock Kernel principal component analysis, in {\em Artificial Neural
  Networks --- ICANN'97\/}, pp. 583--588, Springer Berlin Heidelberg, Berlin,
  Heidelberg.

\bibitem[{Shen} et~al.(2022){Shen}, {Huerta}, {O'Shea}, {Kumar}, \&
  {Zhao}]{Shen-2022}
{Shen}, H., {Huerta}, E.~A., {O'Shea}, E., {Kumar}, P., \& {Zhao}, Z., 2022.
\newblock {Statistically-informed deep learning for gravitational wave
  parameter estimation}, {\it Machine Learning: Science and Technology\/}, {\bf
  3}(1), 015007.

\bibitem[{Slijepcevic} et~al.(2022){Slijepcevic}, {Scaife}, {Walmsley}, \&
  {Bowles}]{Slijepcevic-2022-Neurips}
{Slijepcevic}, I.~V., {Scaife}, A. M.~M., {Walmsley}, M., \& {Bowles}, M.,
  2022.
\newblock {Learning useful representations for radio astronomy ``in the wild''
  with contrastive learning}, {\it arXiv e-prints\/}, p. arXiv:2207.08666.

\bibitem[{Smith} \& {Geach}(2022)]{2022arXiv221103796S}
{Smith}, M.~J. \& {Geach}, J.~E., 2022.
\newblock {Astronomia ex machina: a history, primer, and outlook on neural
  networks in astronomy}, {\it arXiv e-prints\/}, p. arXiv:2211.03796.

\bibitem[Sohl-Dickstein \& Weiss(2015)]{sohl2015deep}
Sohl-Dickstein, J. \& Weiss, E.~A., 2015.
\newblock Deep unsupervised learning using nonequilibrium thermodynamics, in
  {\em International Conference on Machine Learning\/}, pp. 1462--1471.

\bibitem[Song et~al.(2019)Song, Zhang, Gunter, \& Ermon]{song2019generative}
Song, J., Zhang, T., Gunter, C., \& Ermon, S., 2019.
\newblock Generative modeling by estimating gradients of the data distribution,
  in {\em International Conference on Machine Learning\/}, pp. 5711--5720.

\bibitem[{Stein} et~al.(2021){Stein}, {Harrington}, {Blaum}, {Medan}, \&
  {Lukic}]{Stein-2021-Neurips}
{Stein}, G., {Harrington}, P., {Blaum}, J., {Medan}, T., \& {Lukic}, Z., 2021.
\newblock {Self-supervised similarity search for large scientific datasets},
  {\it arXiv e-prints\/}, p. arXiv:2110.13151.

\bibitem[Tenenbaum et~al.(2000)Tenenbaum, de~Silva, \& Langford]{isomap}
Tenenbaum, J.~B., de~Silva, V., \& Langford, J.~C., 2000.
\newblock A global geometric framework for nonlinear dimensionality reduction,
  {\it Science (New York, N.Y.)\/}, {\bf 290(5500)}, 2319–2323.

\bibitem[Tian et~al.(2019)Tian, Krishnan, \& Isola]{CMC}
Tian, Y., Krishnan, D., \& Isola, P., 2019.
\newblock Contrastive multiview coding.

\bibitem[van~der Maaten \& Hinton(2008)]{vanDerMaaten2008}
van~der Maaten, L. \& Hinton, G., 2008.
\newblock Visualizing data using {t-SNE}, {\it Journal of Machine Learning
  Research\/}, {\bf 9}, 2579--2605.

\bibitem[Van Der~Maaten et~al.(2009)Van Der~Maaten, Postma, \& Van~den
  Herik]{van2009dimensionality}
Van Der~Maaten, L., Postma, E., \& Van~den Herik, J., 2009.
\newblock Dimensionality reduction: a comparative review, {\it Journal of
  Machine Learning Research\/}, {\bf 10}(1), 66--71.

\bibitem[{Vega-Ferrero} et~al.(2023){Vega-Ferrero}, {Huertas-Company},
  {Costantin}, {P{\'e}rez-Gonz{\'a}lez}, {Sarmiento}, {Kartaltepe},
  {Pillepich}, {Bagley}, {Finkelstein}, {McGrath}, {Knapen}, {Arrabal Haro},
  {Bell}, {Buitrago}, {Calabr{\`o}}, {Dekel}, {Dickinson}, {Dom{\'\i}nguez
  S{\'a}nchez}, {Elbaz}, {Ferguson}, {Giavalisco}, {Holwerda}, {Kocesvski},
  {Koekemoer}, {Pandya}, {Papovich}, {Pirzkal}, {Primack}, \&
  {Yung}]{2023arXiv230207277V}
{Vega-Ferrero}, J., {Huertas-Company}, M., {Costantin}, L.,
  {P{\'e}rez-Gonz{\'a}lez}, P.~G., {Sarmiento}, R., {Kartaltepe}, J.~S.,
  {Pillepich}, A., {Bagley}, M.~B., {Finkelstein}, S.~L., {McGrath}, E.~J.,
  {Knapen}, J.~H., {Arrabal Haro}, P., {Bell}, E.~F., {Buitrago}, F.,
  {Calabr{\`o}}, A., {Dekel}, A., {Dickinson}, M., {Dom{\'\i}nguez
  S{\'a}nchez}, H., {Elbaz}, D., {Ferguson}, H.~C., {Giavalisco}, M.,
  {Holwerda}, B.~W., {Kocesvski}, D.~D., {Koekemoer}, A.~M., {Pandya}, V.,
  {Papovich}, C., {Pirzkal}, N., {Primack}, J., \& {Yung}, L.~Y.~A., 2023.
\newblock {On the nature of disks at high redshift seen by JWST/CEERS with
  contrastive learning and cosmological simulations}, {\it arXiv e-prints\/},
  p. arXiv:2302.07277.

\bibitem[{Walmsley} et~al.(2022{\natexlab{a}}){Walmsley}, {Scaife}, {Lintott},
  {Lochner}, {Etsebeth}, {G{\'e}ron}, {Dickinson}, {Fortson}, {Kruk},
  {Masters}, {Mantha}, \& {Simmons}]{Walmsley-2022-1}
{Walmsley}, M., {Scaife}, A. M.~M., {Lintott}, C., {Lochner}, M., {Etsebeth},
  V., {G{\'e}ron}, T., {Dickinson}, H., {Fortson}, L., {Kruk}, S., {Masters},
  K.~L., {Mantha}, K.~B., \& {Simmons}, B.~D., 2022{\natexlab{a}}.
\newblock {Practical galaxy morphology tools from deep supervised
  representation learning}, {\it \mnras\/}, {\bf 513}(2), 1581--1599.

\bibitem[{Walmsley} et~al.(2022{\natexlab{b}}){Walmsley}, {Slijepcevic},
  {Bowles}, \& {Scaife}]{Walmsley-2022-2}
{Walmsley}, M., {Slijepcevic}, I.~V., {Bowles}, M., \& {Scaife}, A. M.~M.,
  2022{\natexlab{b}}.
\newblock {Towards Galaxy Foundation Models with Hybrid Contrastive Learning},
  {\it arXiv e-prints\/}, p. arXiv:2206.11927.

\bibitem[{Wei} et~al.(2022){Wei}, {Li}, {Lu}, {Li}, {Liang}, {Dai}, \&
  {Zhang}]{Wei-2022}
{Wei}, S., {Li}, Y., {Lu}, W., {Li}, N., {Liang}, B., {Dai}, W., \& {Zhang},
  Z., 2022.
\newblock {Unsupervised Galaxy Morphological Visual Representation with Deep
  Contrastive Learning}, {\it arXiv e-prints\/}, p. arXiv:2211.07168.

\bibitem[Weinberger \& Saul(2009)]{JMLR-2009}
Weinberger, K.~Q. \& Saul, L.~K., 2009.
\newblock Distance metric learning for large margin nearest neighbor
  classification, {\it Journal of Machine Learning Research\/}, {\bf 10}(9),
  207--244.

\bibitem[{Wright} et~al.(2010){Wright}, {Eisenhardt}, {Mainzer}, {Ressler},
  {Cutri}, {Jarrett}, {Kirkpatrick}, {Padgett}, {McMillan}, {Skrutskie},
  {Stanford}, {Cohen}, {Walker}, {Mather}, {Leisawitz}, {Gautier}, {McLean},
  {Benford}, {Lonsdale}, {Blain}, {Mendez}, {Irace}, {Duval}, {Liu}, {Royer},
  {Heinrichsen}, {Howard}, {Shannon}, {Kendall}, {Walsh}, {Larsen}, {Cardon},
  {Schick}, {Schwalm}, {Abid}, {Fabinsky}, {Naes}, \&
  {Tsai}]{2010AJ....140.1868W}
{Wright}, E.~L., {Eisenhardt}, P. R.~M., {Mainzer}, A.~K., {Ressler}, M.~E.,
  {Cutri}, R.~M., {Jarrett}, T., {Kirkpatrick}, J.~D., {Padgett}, D.,
  {McMillan}, R.~S., {Skrutskie}, M., {Stanford}, S.~A., {Cohen}, M., {Walker},
  R.~G., {Mather}, J.~C., {Leisawitz}, D., {Gautier}, Thomas~N., I., {McLean},
  I., {Benford}, D., {Lonsdale}, C.~J., {Blain}, A., {Mendez}, B., {Irace},
  W.~R., {Duval}, V., {Liu}, F., {Royer}, D., {Heinrichsen}, I., {Howard}, J.,
  {Shannon}, M., {Kendall}, M., {Walsh}, A.~L., {Larsen}, M., {Cardon}, J.~G.,
  {Schick}, S., {Schwalm}, M., {Abid}, M., {Fabinsky}, B., {Naes}, L., \&
  {Tsai}, C.-W., 2010.
\newblock {The Wide-field Infrared Survey Explorer (WISE): Mission Description
  and Initial On-orbit Performance}, {\it \aj\/}, {\bf 140}(6), 1868--1881.

\bibitem[Wu et~al.(2018)Wu, Xiong, Yu, \& Lin]{Wu-2018}
Wu, Z., Xiong, Y., Yu, S., \& Lin, D., 2018.
\newblock Unsupervised feature learning via non-parametric instance-level
  discrimination.

\bibitem[Young(1987)]{MDS}
Young, F., 1987.
\newblock {\it Multidimensional Scaling: History, Theory, and Applications\/}.

\end{thebibliography}








\bsp	
\label{lastpage}
\end{document}